\documentclass[prd,amsmath,floats,amssymb, floatfix,superscriptaddress,nofootinbib,onecolumn]{revtex4} 
\usepackage[plainpages=false, colorlinks=true, anchorcolor=blue, linkcolor=blue, citecolor=blue, bookmarks=false]{hyperref}
\usepackage{graphicx}
\usepackage{multirow}
\usepackage{makecell}

\newcommand{\rthis}[1]{\textcolor{black}{#1}}
\usepackage{appendix}
\usepackage{float}
\usepackage[caption=false]{subfig}
\usepackage[paperwidth=210mm,paperheight=297mm,centering,hmargin=2cm,vmargin=2.5cm]{geometry}

\begin{document}
\include{notations}
\preprint{APS/123-QED}

\title{ Effect of Peak Absolute Magnitude of Type Ia Supernovae and Sound Horizon Values \rthis{on the Hubble Constant} using DESI \rthis{Data Release 1} results}

\author{Shubham Barua}
 \altaffiliation{Email:ph24resch01006@iith.ac.in}
\author{Shantanu Desai}
 \altaffiliation{Email:shntn05@gmail.com}
\affiliation{
 Department of Physics, IIT Hyderabad Kandi, Telangana 502284,  India}




\begin{abstract}

We apply data-motivated priors on the peak absolute magnitude of Type Ia supernovae ($M$) and the sound horizon at the drag epoch ($r_d$), to study how the $M-r_d$ degeneracy affects low redshift measurements of the Hubble constant,  and then compare these estimates  to the Planck estimated value of the Hubble constant. 
We use the data from Pantheon$+$, Cosmic Chronometers, and the Dark Energy Spectroscopic Instrument Data Release 1 (DESI DR1) Baryon Acoustic Oscillations (BAO) results for this purpose. We reaffirm the fact that there is a degeneracy between $M$ and $r_d$, and modifying the $r_d$ values to reconcile the discrepancy in Hubble constant values also requires a change in the peak absolute magnitude $M$. For certain $M$ and $r_d$ priors, the discrepancy is found to reduce to be as low as (1.2-2) $\sigma$ when considering the spatially flat $\Lambda$CDM model. We also notice that for our datasets considered,  the Gaussian prior combination of $M \in \mathcal{N} (-19.253,0.027)$ (obtained from SH0ES) and $r_d \in \mathcal{N} (147.05,0.3)$ Mpc (determined from Planck CMB measurements) is least favored as compared to other prior combinations for the $\Lambda$CDM model.

\end{abstract}


\maketitle
\section{\label{sec:level1}Introduction}
The Hubble constant $H_0$, defined as the Hubble parameter  $H(z = 0)$, is one of the  most important cosmological parameters in the current concordance $\Lambda$CDM model~\cite{freedman}. Considerable efforts have been made to determine its value since Hubble~\cite{hubble} proposed his famous velocity-distance relation~\cite{ChanRatra,Bethapudi,Smoot23}. After more than half a century of efforts,  measurements of the Hubble constant converged to ($72 \pm 8$) km s$^{-1}$Mpc$^{-1}$, obtained from  the Hubble Space Telescope (HST) Key Project~\cite{hst}.

With the advent of precision cosmology, the Hubble constant value came under intense scrutiny. The Planck   collaboration obtained $H_0 = 67.36\pm0.54$ km s$^{-1}$Mpc$^{-1}$ inferred from CMB measurements in the framework of the spatially-flat standard $\Lambda$CDM model~\cite{Planck}. On the other hand, the value of $H_0$ determined from the \textit{Supernovae and $H_0$ for the Equation of State of dark energy} (SH0ES) project, which uses Cepheid-calibrated SNe Ia data is $73.04 \pm 1.04$km s$^{-1}$Mpc$^{-1}$~\cite{shoes}. This difference between the high redshift and low redshift $H_0$ values is known as the Hubble tension~\cite{tensionreview,Verde,DiValentino22}. Various solutions to this tension have been proposed~\cite{hubblehunter} such as early dark energy~\cite{poulin, karwal, poulin_solution, berghaus, KamionkowskiRiess, agrawal,Vagnozzi} and dark energy-dark matter interactions~\cite{Montani} for early universe modifications, while late-time modifications are also viable solutions \cite{raveri,keeley}. It has also been argued that the breakdown in $\Lambda$CDM model implied by the Hubble tension is  a signature of redshift-dependent cosmological parameters~\cite{Colgain24,dainotti_2021, Montani24,montani_2025} (and references therein).
The relation between Hubble tension and other tensions and anomalies in the current concordance model of cosmology can be found in recent reviews~\cite{Periv,Abdalla22,Peebles22,Banik}. 

One of the proposed solutions to fix the Hubble tension conundrum is to modify the value of the sound horizon at the drag epoch $r_d$~\cite{Eisenstein},  which can increase the Hubble expansion rate. The sound horizon is the scale at which the baryons decoupled from the photons during the drag epoch \cite{sunyaev, Peebles}. This serves as a standard ruler in Cosmology known as baryon acoustic oscillations (BAO) \cite{eisensteinspergel, sutherland}. By studying the clustering of galaxies and other cosmic structures, efforts have been made to measure this scale. However, the value of $r_d$ measured from CMB observations is model dependent and is equal to $147\pm0.3$ Mpc \cite{Planck}. 
In the determination of $H_0$, the calibration of $r_d$ becomes important since BAO observations give rise to a strong degeneracy between $H_0$ and $r_d$ in the form of the factor $\frac{c}{r_dH_0}$. Hence, it is evident that model-dependent calibrations can bias the value of $H_0$. 

The discovery of late-time cosmic acceleration was based on Type Ia supernovae observations~\cite{riess1998,Perlmutter,Huterer, weinberg}. The peak absolute magnitude $M$ of SNe Ia plays an important role in constraining cosmological parameters. It appears in the expression for  the cosmic luminosity distance and has a degenerate relation with the Hubble constant $H_0$. To determine the value of $H_0$ from SNe Ia observations, calibration of $M$ becomes important. This is exactly what the SH0ES team did, where using SNe Ia Cepheid hosts they found a value of $M = -19.253\pm0.027$ \cite{shoes}. 
There have been some arguments in the literature that instead of the Hubble tension, one should pay attention to the value of $M$ as it is more fundamental when we think of the distance ladder approach to determine the $H_0$ value~\cite{Efstathiou, Camarena}. This now raises the question of the constancy of $M$ (see, \cite{Perivolaropoulos1, Perivolaropoulos2, Ashall, Evslin, Alestas, denitsa, purba}).  While considering BAO observations with Type Ia SNe measurements, a \rthis{degeneracy} then arises in the $M-r_d$ plane.

\rthis{This work tries to shed further light on this issue by focusing on using the DESI DR1 BAO measurements \cite{DESI} along with the Pantheon$+$ and the Cosmic Chronometer datasets to determine the impact of the values of $r_d$ and $M$ on the inferred Hubble constant values, particularly in relation to the Planck determination of the Hubble constant.}
The outline of this manuscript is as follows. In Section~\ref{sec:level2}, we briefly mention the relevant cosmological relations. Section ~\ref{sec:level3} mentions the datasets and values used in this work, while Section~\ref{sec:level4} describes the approach used. Finally, we present our results and conclusions in Sections \ref{sec:results} and \ref{sec:conclusion}, respectively. 

\section{Cosmological Relations}
\label{sec:level2}
In this work, we consider the spatially-flat $\Lambda$CDM model of the homogeneous and isotropic Universe defined by the FLRW metric~\cite{PDG}: 
$ds^2 = -dt^2 + a(t)^2dR^2$. In this model, the evolution of the Hubble parameter  $H(z)$  is given by:
\begin{equation}
\label{eq:1}
    H(z) = H_0 \sqrt{\Omega_m(1+z)^3 + (1-\Omega_m)},
\end{equation}
where $H_0$ is the Hubble constant and $\Omega_m$ is the dimensionless matter density parameter. For our assumptions, the luminosity distance ($D_L$) is given by 
\begin{equation}
\label{eq:2}
    D_L(z) = c(1+z)\int_0^z \frac{dz'}{H(z')}. 
\end{equation}
For Type Ia supernovae, the relation between its luminosity distance, the apparent magnitude and the peak absolute magnitude is given by~\cite{Huterer} 
\begin{equation}
\label{eq:3}
    m(z) = 5\text{log}_{10}\left[\frac{d_L(z)}{\text{Mpc}}\right]+25+M,
\end{equation}
where $M$ is the peak absolute magnitude and $m(z)$ is the apparent magnitude. 

The DESI DR1 lists the values of $D_M/r_d$, $D_H/r_d$, and $D_V/r_d$ \cite{DESI}. 
By measuring the redshift interval $\Delta z$ along the line-of-sight, we can get an estimate of the Hubble distance at redshift $z$ by 
\begin{equation}
\label{eq:4}
    D_H(z) = \frac{c}{H(z)}.
\end{equation}
The comoving angular diameter distance \cite{augbourg} can be found by measuring the angle $\Delta\theta$ subtended by the BAO feature at a redshift $z$, along the transverse direction and given by:
\begin{equation}
\label{eq:5}
    D_M(z) = (1+z)D_A(z),
\end{equation}
where, $D_A(z)$ is the angular diameter distance \cite{seo} and is related to $D_L$ using the cosmic distance-duality relation~\cite{BoraCDDR} as follows: 
\begin{equation}
\label{eq:6}
    D_A(z) = \frac{D_L(z)}{(1+z)^2}.
\end{equation}
Finally, the BAO measurements also provide an estimation of the spherically averaged distance ($D_V$) given by~\cite{eisensteindv}:
\begin{equation}
\label{eq:7}
    D_V(z) = \left[zD_M(z)^2D_H(z)\right]^{1/3}.
\end{equation}
To make our analysis model independent, we have also done a cosmography analysis using Pad\'e approximants \cite{petreca_2024, aviles_2014, jing_liu_2015, wei_2014, gruber_2014, adachi_2012}.
\rthis{Cosmography \cite{busti_2015, tucker_2005,duncsby_2016} is a model-independent approach to studying the expansion history of the universe in terms of kinematics of the universe. It makes minimal assumptions (large-scale homogeneity and isotropy) while circumventing the need for an expansion model. However, cosmography suffers from the challenge of series expansion convergence \cite{catto_2007, capozziello_2020, gruber_2014,lobo_2020}. Several cosmography  techniques  are based on Taylor series expansions which show convergence problems for high redshifts. To overcome these limitations, an alternative approach based on Pad\'e rational approximant \cite{aviles_2014,mehrabi_2018, wei_2014, zhou_2016a, zhou_2016b, yang_2021, capozzielo_2018, dutta_2018, dutta_2019} has been developed. Using this method, we can study the evolution of the universe at redshifts greater than 1 \cite{agostino_2023}.}
We use the (3,1) Pad\'e approximants and the analytic expressions are stated below (Eqns.~\ref{eq:8} and \ref{eq:9}). We use these expressions in equations \ref{eq:4}, \ref{eq:5},  \ref{eq:6} and \ref{eq:8}  to get the other relevant quantities.

\begin{align}
\label{eq:8}
H^{(3,1)}(z) &= -H_0\Bigg\{ \Big[6(1+z)^2 \left( 4(1+j_0-q_0(1+3q_0)) + (2+5j_0(1+2q_0) - q_0(2+3q_0)(1+5q_0) + s_0)z \right)^2\Big]\Bigg\} \nonumber \\
&\quad \Bigg\{\Big[96(1+j_0-q_0(1+3q_0))^2 + 48(1+j_0-q_0(1+3q_0)) \big( 4+j_0(7+8q_0) - q_0(6+q_0(17+9q_0)) + s_0 \big)z \nonumber \\
&\quad - 6 \Big( 8j_0^3 - j_0^2(49+4q_0(39+23q_0)) - q_0 \big(-56+q_0(-128+q_0(112+q_0(509+462q_0+81q_0^2)))\big) \nonumber \\
&\quad + 2j_0(-34+q_0(-2+q_0(205+q_0(281+78q_0))-6s_0)-9s_0) + 2q_0(10+3q_0(7+q_0))s_0 - s_0^2 - 4(5+3s_0) \Big)z^2  \nonumber \\
&\quad + 2(6+j_0(9+10q_0) - q_0(4+q_0(13+30q_0-9q_0^2) + 3s_0))z^3 \nonumber \\
&\quad+ (2+5j_0(1+2q_0) - q_0(2+3q_0)(1+5q_0) + s_0)(-2+4j_0^2 + j_0(-7+q_0(-23+6q_0)) - 3s_0  \nonumber \\
&\quad+ q_0(4+q_0(13+30q_0-9q_0^2) + 3s_0))z^4\Big]^{-1}\Bigg\}.
\end{align}

\begin{align}
\label{eq:9}
D_L^{(3,1)}(z) &= c H_0^{-1} \Bigg\{ z \Big[ z^2 \big( -4j_0^2 + j_0(q_0(23 - 6q_0) + 7) + q_0(q_0(9q_0^2 - 30q_0 - 13) - 3s_0 - 4) + 3s_0 + 2 \big) \nonumber \\
&\quad + 6z \big( j_0(8q_0 + 7) - q_0(q_0(9q_0 + 17) + 6) + s_0 + 4 \big) + 24(j_0 - q_0(3q_0 + 1) + 1) \Big]\Bigg\} \nonumber \\
&\quad \Bigg\{\Big[ 6 \big( z(5j_0(2q_0 + 1) - q_0(3q_0 + 2)(5q_0 + 1) + s_0 + 2) + 4(j_0 - q_0(3q_0 + 1) + 1) \big) \Big]^{-1} \Bigg\}.
\end{align}
 
\section{Description of Data}
\label{sec:level3}
We describe  the datasets used for our analysis as follows:
\begin{itemize}
    \item For Type Ia supernovae, we use 1590 distinct samples from the Pantheon$+$ compilation~\cite{scolnic} in the redshift range 0.001 to 2.26 \footnote{The data release can be found at https://github.com/PantheonPlusSH0ES/DataRelease}. All the uncertainties have been incorporated in the covariance matrix provided along with the dataset. 
    \item The cosmic chronometer~\cite{Jimenez} dataset has been obtained from~\cite{moresco1, moresco2, moresco3, ratsimbazafy,  stern, borghi, joan, cong} in the redshift range $0.07 \le z \le 1.965$. We use the covariance matrix for computations as described in \cite{moresco4}. We tabulate the $H(z)$ values used for the analysis in Table~\ref{table1}, which can also be found in Table (1.1) of~\cite{Moresco_2307} The last 15 $H(z)$ measurements of Table \ref{table1} are correlated~\footnote{https://gitlab.com/mmoresco/CCcovariance}.
    \item DESI DR1 listed in Table 1 of \cite{DESI}. We consider both isotropic and anisotropic BAO data that include the observables-BGS, LRG, ELG and QSOs. We also incorporated  the correlation coefficients ($r$) listed in the table. The redshift range  of this sample is between 0.1 and 4.16. 
\end{itemize}
To fix the absolute value of the peak magnitude and the sound horizon, we use four  data-motivated priors obtained from  literature as follows:
\begin{itemize}
    \item $M = -19.253 \pm 0.027$ obtained by the Cepheid calibration of Type Ia SNe based on SH0ES observations~\cite{shoes}.
    \item $M = -19.362^{+0.078}_{-0.067}$ considering a model-independent method~\cite{36} using Type Ia SNe Ia observations along with BAO and CC data.
    \item $M = -19.396 \pm 0.015$. This value has been obtained using Gaussian Process Regression~\cite{Seikel} for a model-independent and non-parametric approach, similar to the analysis in~\cite{Dinda}.
    \item $M = -19.401 \pm 0.027$ obtained by using a 
    model-independent binning technique, which combined type Ia SNe observations with anisotropic BAO observations~\cite{7}.
    \item $M = -19.420 \pm 0.014$ where $\Lambda$CDM model was used to calibrate the Type Ia SNe with Planck CMB data \cite{28Greene}.
\end{itemize}
For the sound horizon values we used the following priors from the literature:
\begin{itemize}
    \item $r_d = 137\pm4.5$ Mpc using the angular diameter distances to three time-delay lenses from the H0LiCOW collaboration in a model-independent approach~\cite{wojtak}.
    \item $r_d = 139.7^{+5.2}_{-4.5}$ Mpc obtained by using BAO observations and gravitationally time-delay lensed quasars from H0LiCOW observations using a model-independent approach~\cite{liusoundhorizon}.
    \item $r_d = 147.05\pm0.3$ Mpc which is the value obtained by the Planck collaboration \cite{Planck} \rthis{(TT$+$TE$+$EE$+$low E)}.
    \item $r_d = 148\pm3.6$ Mpc using the model-independent polynomial expansions approach \cite{zhang}.
\end{itemize}

For our analysis, when we use all of the above datasets together, we only consider data points which span the same range of redshifts. Therefore, we work with the redshift range from 0.1 - 1.965, the lower limit coming from the redshift range in the DESI data, while the upper limit corresponds to the highest value in  the Cosmic Chronometer (CC) dataset. The CC data is utilized in this work to tighten the constraints on $H_0$ compared to using only BAO and SNe Ia data, when applying uniform priors on $M$ and $r_d$.

\begin{table}[htbp!]
\caption{32 $H(z)$ data}
\label{table1}
\centering
    \begin{tabular}{|c|c|c|}
        \hline
        \thead{\boldmath$z$} & \thead{\boldmath$H_0$ (km/s/Mpc)} & \thead{Reference}\\
        \hline
        $0.07$   & $69.0\pm19.6$  & \cite{cong} \\
        $0.09$   & $69.0\pm12.0$  & \cite{joan} \\
        $0.12$   & $68.6\pm26.2$  & \cite{cong} \\
        $0.17$   & $83.0\pm8.0$   & \cite{joan} \\
        $0.2$    & $72.9\pm29.6$  & \cite{cong} \\
        $0.27$   & $77.0\pm14.0$  & \cite{joan} \\
        $0.28$   & $88.8\pm36.6$  & \cite{cong} \\
        $0.4$    & $95.0\pm17.0$  & \cite{joan} \\
        $0.47$   & $89.0\pm50.0$  & \cite{ratsimbazafy} \\
        $0.48$   & $97.0\pm62.0$  & \cite{stern} \\
        $0.75$   & $98.8\pm33.6$  & \cite{borghi} \\
        $0.88$   & $90.0\pm40.0$  & \cite{stern} \\
        $0.9$    & $117.0\pm23.0$ & \cite{joan} \\
        $1.3$    & $168.0\pm17.0$ & \cite{joan} \\
        $1.43$   & $177.0\pm18.0$ & \cite{joan} \\
        $1.53$   & $140.0\pm14.0$ & \cite{joan} \\
        $1.75$   & $202.0\pm40.0$ & \cite{joan} \\
        $0.1791$ & $74.91$        & \cite{moresco4} \\
        $0.1993$ & $74.96$        & \cite{moresco4} \\
        $0.3519$ & $82.78$        & \cite{moresco4} \\
        $0.3802$ & $83.0$         & \cite{moresco4} \\
        $0.4004$ & $76.97$        & \cite{moresco4} \\
        $0.4247$ & $87.08$        & \cite{moresco4} \\
        $0.4497$ & $92.78$        & \cite{moresco4} \\
        $0.4783$ & $80.91$        & \cite{moresco4} \\
        $0.5929$ & $103.8$        & \cite{moresco4} \\
        $0.6797$ & $91.6$         & \cite{moresco4} \\
        $0.7812$ & $104.5$        & \cite{moresco4} \\
        $0.8754$ & $125.1$        & \cite{moresco4} \\
        $1.037$  & $153.7$        & \cite{moresco4} \\
        $1.363$  & $160.0$        & \cite{moresco4} \\
        $1.965$  & $186.5$        & \cite{moresco4} \\
        \hline
    \end{tabular}
\end{table}

\section{Methodology}
\label{sec:level4}
The parameters $\{H_0, \Omega_m, r_d, M\}$ are constrained using Bayesian inference. For this purpose, the posteriors are sampled using  NAUTILUS~\cite{nautilus}, while the marginalized posteriors were  generated using \texttt{getdist}\cite{getdist}.

First, using Pantheon$+$, DESI BAO, and CC unbinned data points in the common redshift range 0.1 - 1.965, we constrain $H_0, \Omega_m, M$, and $r_d$ values using uniform priors for the parameters. Subsequently, we apply a Gaussian prior on either $M$ or $r_d$ using one of the values described in Section~\ref{sec:level3} and assign a uniform prior on the other. Finally, we apply Gaussian priors on both $M$ and $r_d$.
As noted in \cite{colgain}, one DESI LRG datum at $z_\text{eff} = 0.51$ has been identified as a potential outlier. To account for this anomalous data point, we conducted our analysis, both including and excluding this data point. However, we find no significant differences between the two results. Therefore, all the results shown in this work include this data point.
Finally, we compare the $H_0$ values obtained from our analyses to the Planck cosmological analyses (obtained from TT,TE,EE$+$low E$+$lensing), which has the value $67.36\pm0.54$ km s$^{-1}$ Mpc$^{-1}$~\cite{Planck} and quantify the significance of the \rthis{discrepancy}. It has been known for a while that cosmological parameters inferred through Bayesian analysis could be prior dependent~\cite{Patel}.
Additionally, as has been pointed out in \cite{Chen_2024}, the choice of priors on the peak absolute magnitude plays a pivotal role in the analysis of the Pantheon$+$ dataset. Therefore, we further sub-divide our analysis into two parts, wherein we use both Gaussian and uniform priors on $r_d$ and $M$. For Gaussian priors, we use the values mentioned in Section~\ref{sec:level3}, while for uniform priors, we use $r_d \in \mathcal{U} (0, 200) \text{ and } M \in \mathcal{U} (-21, -18)$. Additionally, uniform priors have been used for $H_0$ and $\Omega_m$ as given by:
\begin{equation*}
    H_0 \in \mathcal{U} (10, 200) \text{ and } \Omega_m \in \mathcal{U} (0,1).
\end{equation*}

For  this analysis, the total likelihood is given by:
\begin{equation}
    L(\mathbf{\theta}) \propto e^{-\chi^2 / 2},
\end{equation}
where $\theta$ represents the parameter vector (or subset depending on which parameters are fixed)$\{H_0, \Omega_m, r_d, M\}$ and 
\begin{equation}
    \chi^2_\text{tot} = \chi^2_\text{SNe} + \chi^2_\text{BAO} + \chi^2_\text{CC}.
\end{equation}
The likelihood construction for the SNe Ia datset can be found in \cite{pantheon} and for the CC dataset can be found in \cite{moresco}. For the BAO data, we construct a covariance matrix utilizing the correlation factors and the standard deviations and consider a Gaussian likelihood \cite{DESI, ruchika_2024}.
We have also conducted the same analysis without including CC data, as presented in Appendix \ref{appA}, to demonstrate that the inclusion of CC data leads to more stringent constraints on $H_0$ when uniform priors are applied to $M$ and $r_d$. 
We also do a cosmography based analysis. The analytic expressions for the luminosity distance and the Hubble constant for the same can be found in Eqns.~\ref{eq:8} and \ref{eq:9}.
We use Bayesian model comparison  and calculate the Bayes factor~\cite{trotta_2017,Krishak20} in order to determine the preferred prior combination on $M$ and $r_d$.  In this case, we have the same underlying theoretical model, namely $\Lambda$CDM and the same data. Therefore, the Bayes factor provides a measure of the relative efficacy of the priors for the given dataset and $\Lambda$CDM model. The null hypothesis in this case is the scenario in which we apply SH0ES prior ($M \in \mathcal{N}$ (-19.253,0.027)to $M$ and the CMB inferred value of  ($r_d \in \mathcal{N} (147.05,0.3)$ Mpc) of $r_d$ in the spatially flat $\Lambda$ CDM model (Table \ref{table8}). Henceforth, we will refer to this prior combination as the standard prior combination.
To determine the significance from the Bayes factors, we use the  Jeffreys scale~\cite{trotta_2017}.
We now present  our results for our analyses in the next section.

\section{Results}
\label{sec:results}
\begin{figure}
    \centering
    \includegraphics[width=0.9\linewidth]{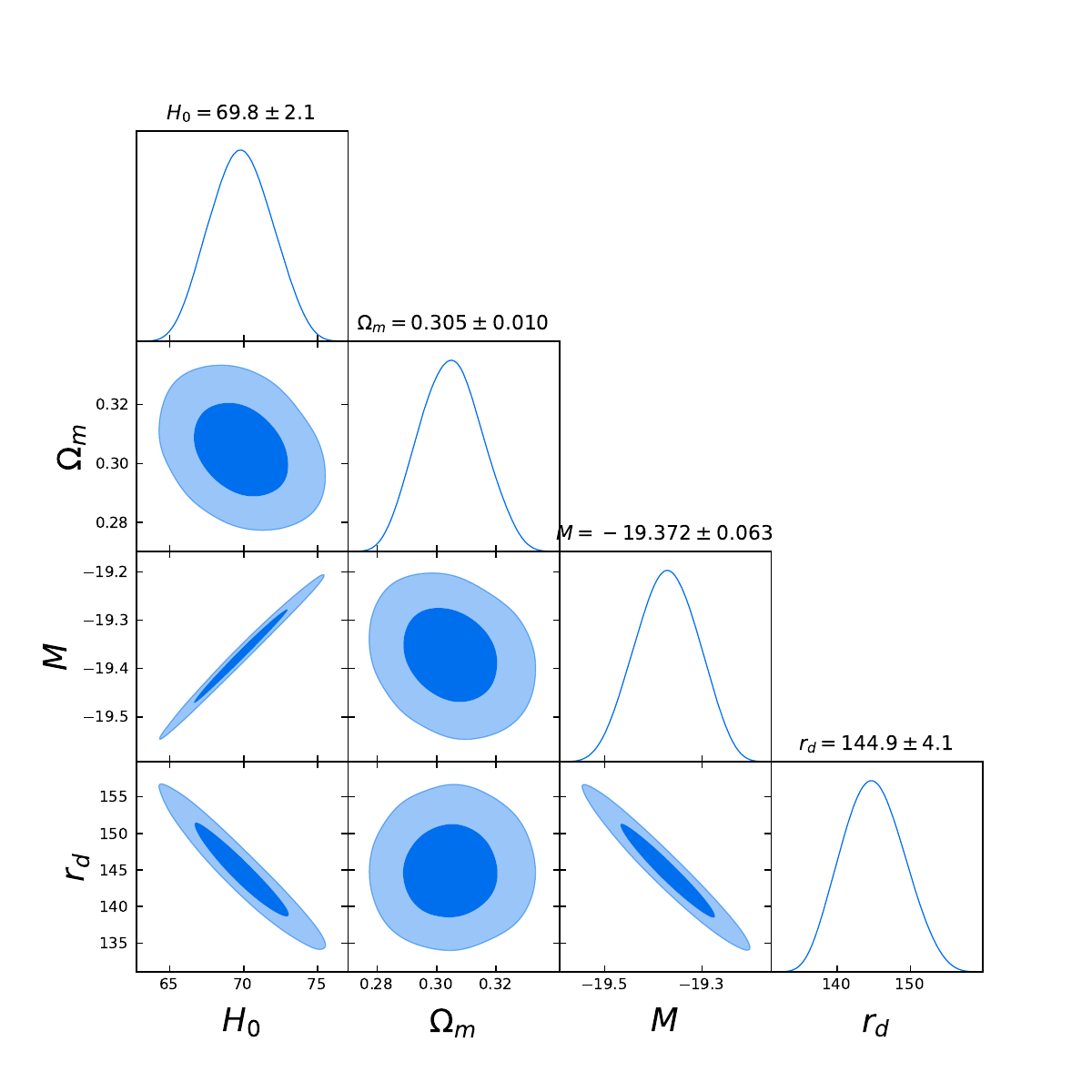}
    \caption{Pantheon Plus, DESI and CC dataset with $H_0$, $\Omega_m$, $r_d$ and $M$ as parameters (uniform priors). The contours represent marginalized 68$\%$ and 99\% credible intervals.}
    \label{fig1}
\end{figure}

Our results \rthis{for the spatially-flat $\Lambda$CDM case} with all the aforementioned analyses can be found in Tables \ref{table2}, \ref{table4}, \ref{table6}, \ref{table8}, and Figure \ref{fig1}. We state the results obtained using a cosmography-based approach with  Pad\'e approximants \cite{petreca_2024, aviles_2014, jing_liu_2015, wei_2014, gruber_2014, adachi_2012} in Tables~\ref{table3}, \ref{table5}, \ref{table7}, and \ref{table9}.

\begin{enumerate}
    \item Uniform priors on $r_d$ and $M$: (Figure~\ref{fig1}, Table~\ref{table2} and \ref{table3})
    \begin{itemize}
        \item  We see that the Pantheon$+$, DESI and CC datasets favor $H_0 = 69.8 \pm 2.1$ km s$^{-1}$ Mpc$^{-1}$ for $M = -19.37 \pm 0.06$ and a value of $r_d = 144.9\pm4.1$ Mpc for uniform priors on both $r_d$ and $M$. The  corresponding  Bayes' factor is equal  to 70, implying that this is very strongly favored compared to the standard prior combination.
        \item \rthis{Cosmography approach gives a value of $H_0 = 68.8\pm2.6$ km s$^{-1}$ Mpc$^{-1}$ and similar constraints on $M$ and $r_d$. The Bayes' factor value is 71.5. This result is similar to what we got using standard $\Lambda$CDM model.}
        \item \rthis{The discrepancy in the cosmography case is reduced to as low as $0.54\sigma$ in comparison to $1.13\sigma$ in the standard $\Lambda$CDM case.}
     \end{itemize}
     The values of $H_0$ \rthis{(for the model-dependent case)} are consistent with the Planck value within $1.2\sigma$, and is also  in agreement (within $1\sigma$) with that obtained from a  joint analysis of CC $+$ BAO $+$ Pantheon$+$  quasar angular size  $+$ Mg II and  CIV quasar measurements $+$ GRB data, viz. $H_0 = 69.25\pm 2.42 $ km s$^{-1}$ Mpc$^{-1}$~\cite{caoratra}.
    \item Uniform priors on $r_d$ and Gaussian priors on $M$: (Table~\ref{table4} and \ref{table5})
    \begin{itemize}
        \item When applying Gaussian priors to $M$, the mean values of $r_d$ increase as $M$ decreases. This reduces the \rthis{discrepancy} to as much as  $1.2 \sigma$ for $M \in \mathcal{N} (-19.42,0.014)$ and $r_d = 147.7\pm1.6$ Mpc.
        \item \rthis{A cosmography-based analysis yields comparable values of $H_0$. However, the central $H_0$ values are marginally lower than those obtained from the corresponding prior combination in the standard $\Lambda$CDM framework.}
        \item \rthis{The discrepancies are considerably lower in the cosmography approach as compared to the standard $\Lambda$CDM case and goes as low as $0.2\sigma$.}
    \end{itemize}
    \item Uniform priors on $M$ and Gaussian priors on $r_d$: (Table~\ref{table6} and \ref{table7})
    \begin{itemize}
        \item This follows a similar trend as the previous case,  where $M$ decreases with increasing values of  $r_d$.
        \item We find that the $H_0$ is correlated with $M$ and decreasing the $M$ values also reduces the estimate of $H_0$,  with the \rthis{discrepancy in the Hubble constant values} ranging from (0.97-2.42)$\sigma$.
        \item \rthis{The results from cosmography-based analysis support those obtained using the $\Lambda$CDM model. However, the inferred $H_0$ values remain slightly lower than those obtained within the $\Lambda$CDM framework for the same prior combination.}
        \item \rthis{The Hubble tension in the cosmography case ranges between $(0.3-1.84)\sigma$.}
    \end{itemize}
    \item Gaussian priors on both $r_d$ and $M$ (Table~\ref{table8} and \ref{table9}). Here, we considered twenty different use-cases. Our conclusions are as follows:
    \begin{itemize}
        \item For SH0ES prior on $M$, the \rthis{discrepancy} with the Planck value remains high ($\sim 5 \sigma$ \rthis{for $\Lambda$CDM case and $\sim 4 \sigma$ for cosmography approach}), independent of  the change in the value of $r_d$.
        \item For priors on $M$ other than the SH0ES value, the tension reduces considerably to  $2\sigma$, and it keeps on decreasing as the value of $M$ decreases and $r_d$ increases. For $M \in \mathcal{N} (-19.42,0.014)$ and $r_d \in \mathcal{N} (148,3.6)$ Mpc, the discrepancy becomes only $1.2 \sigma$.
        \item For a fixed value of  $r_d$, the \rthis{discrepancy} decreases as $M$ decreases. However, note the very gradual decrease in the mean Hubble constant value from 69.44 km s$^{-1}$ Mpc$^{-1}$ to 68.88 km s$^{-1}$ Mpc$^{-1}$ for $M \in \mathcal{N} (-19.401,0.027)$, when $r_d$ is increased. There is a similar trend for other $M$ priors as well.
        \item \rthis{Similar trends as in the standard $\Lambda$CDM case can be observed in the cosmography approach as well. However, the discrepancy with the Planck $H_0$ value is way lower than in the $\Lambda$CDM case.}
    \end{itemize}
    \item Comparison of priors based on Bayes Factors:
    \begin{itemize}
        \item Tables~\ref{table8} \rthis{and \ref{table9}} demonstrates that all other prior combinations are decisively favored compared to the standard prior combination for the spatially-flat $\Lambda$CDM model.
    \end{itemize}
\end{enumerate}

The $H_0$ values which we get for all priors on  $M$ except  $-19.253\pm0.027$, are consistent  with the  $H_0$ value  of $69.03\pm1.75$ km s$^{-1}$ Mpc$^{-1}$, obtained using  TRGB and JAGB methods with JWST data~\cite{Freedman_2024} to within 
$\sim 1\sigma$ maximum.
It is interesting to note that this consistency occurs when we consider low values of $M$. 
Further, our $H_0$ values determined for $M$ other than the SH0ES prior are very much in agreement with \cite{recentdesi}, which circumvented calibrations related to the sound horizon at the baryon drag epoch or $M$ of Type Ia SNe and so is a purely data-driven method. 
\rthis{We note that in all cases, the cosmography approach gives us a smaller discrepancy  with the  Planck value when compared to the $\Lambda$CDM case. This is due to a combination of lower central $H_0$ values (in comparison with the $\Lambda$CDM model inferred values) and large error bars.} 

\begin{table}
\caption{\rthis{Discrepancy} in $H_0$ estimate compared to that from Planck Cosmology~\cite{Planck} for a uniform prior on $M$ and a Uniform prior on $r_d \in (0, 200)$. The standard prior combination ($M \in \mathcal{N} (-19.253,0.027)$) and $r_d \in \mathcal{N} (147.05,0.3)$ Mpc) in Table \ref{table8} has Bayes' Factor of 1.}
\label{table2}
\centering
    \begin{tabular}{|c|c|c|c|c|c|}
        \hline
        \thead{\boldmath$M$} & \thead{\boldmath$r_d$ (Mpc)} & \thead{\boldmath$H_0$ (km/s/Mpc)} & \thead{\boldmath$\Omega_m$} & \thead{Bayes' Factor} & \thead{Discrepancy (in $\sigma)$}\\
        \hline
        $-19.37\pm0.06$ & $144.9\pm4.1$ & $69.8\pm2.1$ & $0.305\pm0.01$ & 70 & 1.13 \\
        \hline
    \end{tabular}
\end{table}

\begin{table}[htbp!]
\caption{Cosmographic Parameters for a uniform prior on $M$ and a uniform prior on $r_d \in (0, 200)$. The standard prior combination (cf. Table~\ref{table2}) in Table \ref{table9} has Bayes' Factor of 1.}
\label{table3}
\centering
    \begin{tabular}{|c|c|c|c|c|c|c|c|}
        \hline
        \thead{\boldmath$M$} & \thead{\boldmath$r_d$ (Mpc)} & \thead{\boldmath$H_0$ (km/s/Mpc)} & \thead{\boldmath$q_0$} & \thead{\boldmath$j_0$} & \thead{\boldmath$s_0$} & \thead{Bayes' Factor} & \thead{Discrepancy (in $\sigma)$}\\
        \hline
        $-19.38\pm0.08$ & $145.2^{+4.8}_{-5.4}$ & $68.8\pm2.6$ & $-0.426\pm0.09$ & $0.74^{+0.36}_{-0.47}$ & $0.58^{+0.38}_{-1.6}$ &  71.52 & 0.54 \\
        \hline
    \end{tabular}
\end{table}

\begin{table}[htbp!]
\caption{\rthis{Discrepancy} in $H_0$ estimate compared to that from Planck Cosmology~\cite{Planck} for a Gaussian prior on $M$ and a uniform prior on $r_d \in (50, 200)$. The standard prior combination (cf. Table~\ref{table2}) in Table \ref{table8} has Bayes' Factor of 1.}
\label{table4}
\centering
    \begin{tabular}{|c|c|c|c|c|c|}
        \hline
        \thead{\boldmath$M$} & \thead{\boldmath$r_d$ (Mpc)} & \thead{\boldmath$H_0$ (km/s/Mpc)} & \thead{\boldmath$\Omega_m$} & \thead{Bayes' Factor} & \thead{Discrepancy (in $\sigma)$}\\
        \hline
        $-19.253\pm0.027$ & $138.3\pm1.9$ & $73.31\pm0.90$ & $0.301\pm0.012$ & 354 & 5.63 \\
        $-19.362\pm0.072$ & $144.6\pm3.5$ & $70.0\pm1.7$   & $0.305\pm0.012$ & 804 & 1.48 \\
        $-19.396\pm0.015$ & $146.3\pm1.6$ & $69.06\pm0.56$ & $0.306\pm0.012$ & 1012 & 2.23 \\
        $-19.401\pm0.027$ & $146.5\pm2.1$ & $68.99\pm0.86$ & $0.306\pm0.012$ & 972 & 1.6 \\
        $-19.420\pm0.014$ & $147.7\pm1.6$ & $68.31\pm0.52$ & $0.307\pm0.012$ & 880 & 1.25 \\
        \hline
    \end{tabular}
\end{table}

\begin{table}[htbp!]
\caption{Cosmographic Parameters for a Gaussian prior on $M$ and a uniform prior on $r_d \in (0, 200)$. The standard prior combination (cf. Table~\ref{table2}) in Table \ref{table9} has Bayes' Factor of  1.}
\label{table5}
\centering
    \begin{tabular}{|c|c|c|c|c|c|c|c|}
        \hline
        \thead{\boldmath$M$} & \thead{\boldmath$r_d$ (Mpc)} & \thead{\boldmath$H_0$ (km/s/Mpc)} & \thead{\boldmath$q_0$} & \thead{\boldmath$j_0$} & \thead{\boldmath$s_0$} & \thead{Bayes' Factor} & \thead{Discrepancy (in $\sigma)$}\\
        \hline
        $-19.253\pm0.027$ & $138.1\pm1.9$ & $72.5\pm1.0$ & $-0.431\pm0.088$ & $0.71^{+0.35}_{-0.45}$ & $0.40^{+0.37}_{-1.5}$ & 330.3 & 4.52 \\
        $-19.362\pm0.072$ & $144.5\pm3.5$ & $69.8\pm1.8$ & $-0.425\pm0.091$ & $0.73^{+0.36}_{-0.47}$ & $0.55^{+0.38}_{-1.6}$ & 804.3 & 1.3\\
        $-19.396\pm0.015$ & $146.1\pm1.6$ & $68.29\pm0.76$ & $-0.424\pm0.092$ & $0.74^{+0.36}_{-0.49}$ & $0.59^{+0.38}_{-1.6}$ & 1043.2 & 1\\
        $-19.401\pm0.027$ & $146.3\pm2.1$ & $68.2\pm1.0$ & $-0.424\pm0.093$ & $0.74^{+0.36}_{-0.48}$ & $0.61^{+0.37}_{-1.6}$ & 992.3 & 0.74\\
        $-19.420\pm0.014$ & $147.5\pm1.6$ & $67.54\pm0.73$ & $-0.422\pm0.091$ & $0.74^{+0.36}_{-0.48}$ & $0.61^{+0.39}_{-1.6}$ & 925.2 & 0.2\\
        \hline
    \end{tabular}
\end{table}

\begin{table}[htbp!]
\caption{\rthis{Discrepancy} in $H_0$ estimate compared to that from Planck Cosmology~\cite{Planck} for a Gaussian prior on $r_d$ and uniform prior on $M \in (-21, -18)$. The standard prior combination (cf. Table~\ref{table2}) in Table \ref{table8} has Bayes' Factor of 1.}
\label{table6}
\centering
    \begin{tabular}{|c|c|c|c|c|c|}
        \hline
        \thead{\boldmath$r_d$ (Mpc)} & \thead{\boldmath$M$} & \thead{\boldmath$H_0$ (km/s/Mpc)} & \thead{\boldmath$\Omega_m$} & \thead{Bayes' Factor} & \thead{Discrepancy (in $\sigma)$}\\
        \hline
        $137.00\pm4.5$  & $-19.308\pm0.052$ & $71.9\pm1.8$   & $0.305\pm0.013$  & 317 & 2.42 \\
        $139.70\pm4.85$ & $-19.331\pm0.055$ & $71.1\pm1.9$   & $0.304\pm0.013$  & 464 & 1.94 \\
        $147.05\pm0.3$  & $-19.404\pm0.02$  & $68.78\pm0.76$ & $0.305\pm0.012$  & 706 & 1.57 \\
        $148.00\pm3.6$  & $-19.402\pm0.048$ & $68.8\pm1.6$   & $0.305\pm0.0143$ & 590 & 0.97 \\
        \hline
    \end{tabular}
\end{table}

\begin{table}[htbp!]
\caption{Cosmographic Parameters for a Gaussian prior on $r_d$ and uniform prior on $M \in (-21, -18)$. The standard prior combination (cf. Table~\ref{table2}) in Table \ref{table9} has Bayes' Factor of 1.}
\label{table7}
\centering
    \begin{tabular}{|c|c|c|c|c|c|c|c|}
        \hline
        \thead{\boldmath$r_d$ (Mpc)} & \thead{\boldmath$M$} & \thead{\boldmath$H_0$ (km/s/Mpc)} & \thead{\boldmath$q_0$} & \thead{\boldmath$j_0$} & \thead{\boldmath$s_0$} & \thead{Bayes' Factor} & \thead{Discrepancy (in $\sigma)$}\\
        \hline
        $137.00\pm4.5$ & $-19.311\pm0.053$ & $71.0\pm1.9$ & $-0.419\pm0.090$ & $0.69^{+0.35}_{-0.46}$ & $0.39^{+0.35}_{-1.5}$ & 415.7 & 1.84\\
        $139.70\pm4.85$ & $-19.337\pm0.055$ & $70.2\pm1.9$ & $-0.421\pm0.091$ & $0.70^{+0.36}_{-0.46}$ & $0.44^{+0.35}_{-1.5}$ & 620.2 & 1.44\\
        $147.05\pm0.3$  & $-19.409\pm0.02$ & $67.9\pm1.0$ & $-0.427\pm0.091$ & $0.75^{+0.36}_{-0.48}$ & $0.63^{+0.40}_{-1.6}$ & 1012.3 & 0.48\\
        $148.00\pm3.6$  & $-19.407\pm0.048$ & $67.9\pm1.7$ & $-0.427\pm0.092$ & $0.75^{+0.37}_{-0.48}$ & $0.64^{+0.40}_{-1.6}$ & 804.3 & 0.3\\
        \hline
    \end{tabular}
\end{table}

\begin{table}[htbp!]
\caption{\rthis{Discrepancy} in $H_0$ estimate compared to that from Planck Cosmology~\cite{Planck}  for a  Gaussian prior on $r_d$ and $M$. For the standard prior combination (cf. Table~\ref{table2}), the Bayes' Factor is 1, which is the null hypothesis.}
\label{table8}
\centering
    \begin{tabular}{|c|c|c|c|c|c|c|}
        \hline
        \thead{\boldmath$M$ prior} & \thead{\boldmath$r_d$ (Mpc)} & \thead{\boldmath$H_0$ (km/s/Mpc)} & \thead{\boldmath$\Omega_m$} & \thead{Bayes' Factor} & \thead{Discrepancy (in $\sigma)$}\\
        \hline
        \multirow{4}{8em}{$-19.253\pm0.027$} & $137\pm4.5$ & $73.39\pm0.88$ & $0.301\pm0.012$ & $4.2\times10^3$ & 5.84 \\
        & $139.7\pm4.85$ & $73.29\pm0.88$ & $0.300\pm0.012$ & $3.9\times10^3$ & 5.74 \\
        & $147.05\pm0.3$ & $70.71\pm0.63$ & $0.280\pm0.010$ & 1 & 4.03 \\
        & $148\pm3.6$    & $72.57\pm0.87$ & $0.295\pm0.012$ & 320 & 5.08 \\
        \hline
        \multirow{4}{8em}{$-19.362\pm0.072$} & $137\pm4.5$ & $71.3\pm1.4$ & $0.306\pm0.013$ & $3.5\times10^3$ & 2.62 \\
        & $139.7\pm4.85$ & $70.8\pm1.5$   & $0.305\pm0.012$ & $5.8\times10^3$ & 2.15 \\
        & $147.05\pm0.3$ & $68.85\pm0.72$ & $0.304\pm0.013$ & $10^4$ & 1.65 \\
        & $148\pm3.6$    & $69.3\pm1.3$   & $0.304\pm0.012$ & $7.3\times10^3$ & 1.38 \\
        \hline
        \multirow{4}{8em}{$-19.396\pm0.015$} & $137\pm4.5$ & $69.21\pm0.54$ & $0.311\pm0.012$ & $1.9\times10^3$ & 2.42 \\
        & $139.7\pm4.85$  & $69.16\pm0.56$ & $0.308\pm0.012$ & $5.1\times10^3$ & 2.31 \\
        & $147.05\pm0.3$  & $68.95\pm0.51$ & $0.303\pm0.010$ & $33\times10^3$ & 2.15 \\
        & $148\pm3.6$     & $69.02\pm0.55$ & $0.304\pm0.012$ & $13.9\times10^3$ & 2.15 \\
        \hline
        \multirow{4}{8em}{$-19.401\pm0.027$} & $137\pm4.5$ & $69.44\pm0.83$ & $0.310\pm0.013$ & $1.8\times10^3$ & 2.1 \\
        & $139.7\pm4.85$ & $69.31\pm0.85$ & $0.308\pm0.012$ & $4.8\times10^3$ & 1.93 \\
        & $147.05\pm0.3$ & $68.80\pm0.64$ & $0.305\pm0.011$ & $26\times10^3$ & 1.72 \\
        & $148\pm3.6$    & $68.88\pm0.81$ & $0.305\pm0.012$ & $12.8\times10^3$ & 1.56 \\
        \hline
        \multirow{4}{8em}{$-19.420\pm0.014$} & $137\pm4.5$ & $68.46\pm0.52$ & $0.313\pm0.012$ & $837$ & 1.47 \\
        & $139.7\pm4.85$ & $68.41\pm0.54$ & $0.310\pm0.012$ & $2.9\times10^3$ & 1.38 \\
        & $147.05\pm0.3$ & $68.39\pm0.48$ & $0.311\pm0.098$ & $29\times10^3$ & 1.43 \\
        & $148\pm3.6$    & $68.31\pm0.52$ & $0.306\pm0.012$ & $13.5\times10^3$ & 1.27 \\
        \hline
    \end{tabular}
\end{table}

\begin{table}[htbp!]
\caption{Cosmographic parameters for a  Gaussian prior on $r_d$ and $M$. For the standard prior combination (cf. Table~\ref{table2}), the Bayes' Factor is 1, which is the null hypothesis.}
\label{table9}
\centering
    \begin{tabular}{|c|c|c|c|c|c|c|c|}
        \hline
        \thead{\boldmath$M$ prior} & \thead{\boldmath$r_d$ (Mpc)} & \thead{\boldmath$H_0$ (km/s/Mpc)} & \thead{\boldmath$q_0$} & \thead{\boldmath$j_0$} & \thead{\boldmath$s_0$} & \thead{Bayes' Factor} & \thead{Discrepancy (in $\sigma)$}\\
        \hline
        \multirow{4}{8em}{$-19.253\pm0.027$} & $137\pm4.5$ & $72.5\pm1.0$ & $-0.428\pm0.089$ & $0.70^{+0.35}_{-0.44}$ & $0.37^{+0.37}_{-1.4}$ & $5.2\times10^3$ & 4.52\\
        & $139.7\pm4.85$ & $72.4\pm1.0$ & $-0.432\pm0.088$ & $0.71^{+0.35}_{-0.44}$ & $0.39^{+0.37}_{-1.5}$ & $4.9\times10^3$ & 4.43\\
        & $147.05\pm0.3$ & $70.26\pm0.86$ & $-0.522\pm0.08$ & $0.99^{+0.34}_{-0.45}$ & $1.27^{+0.67}_{-2.0}$ & 1 & 2.86\\
        & $148\pm3.6$    & $71.9\pm1.0$ & $-0.456\pm0.087$ & $0.79^{+0.36}_{-0.45}$ & $0.62^{+0.46}_{-1.6}$ & 357.8 & 3.9\\
        \hline
        \multirow{4}{8em}{$-19.362\pm0.072$} & $137\pm4.5$ & $70.4\pm1.5$ & $-0.415\pm0.091$ & $0.68^{+0.35}_{-0.45}$ & $0.39^{+0.34}_{-1.5}$ & $4.8\times10^3$ & 1.9\\
        & $139.7\pm4.85$ & $69.8\pm1.6$ & $-0.419\pm0.090$ & $0.70^{+0.36}_{-0.46}$ & $0.44^{+0.38}_{-1.5}$ & $7.7\times10^3$ & 1.44\\
        & $147.05\pm0.3$ & $68.01\pm0.96$ & $-0.432\pm0.089$ & $0.76^{+0.36}_{-0.47}$ & $0.63^{+0.43}_{-1.6}$ & $1.3\times10^4$ & 0.6\\
        & $148\pm3.6$    & $68.4\pm1.4$ & $-0.431\pm0.092$ & $0.75^{+0.36}_{-0.48}$ & $0.64^{+0.30}_{-1.7}$ & $9.6\times10^3$ & 0.7\\
        \hline
        \multirow{4}{8em}{$-19.396\pm0.015$} & $137\pm4.5$ & $68.36\pm0.76$ & $-0.407\pm0.092$ & $0.69^{+0.36}_{-0.48}$ & $0.47^{+0.34}_{-1.5}$ & $2.9\times10^3$ & 1.07\\
        & $139.7\pm4.85$ & $68.33\pm0.77$ & $-0.413\pm0.093$ & $0.71^{+0.36}_{-0.49}$ & $0.53^{+0.36}_{-1.6}$ & $7.3\times10^3$ & 1.03\\
        & $147.05\pm0.3$ & $68.20\pm0.74$ & $-0.44\pm0.084$ & $0.78^{+0.36}_{-0.46}$ & $0.67^{+0.46}_{-1.7}$ & $4.1\times10^4$ & 0.92\\
        & $148\pm3.6$    & $68.26\pm0.76$ & $-0.428\pm0.090$ & $0.74^{+0.36}_{-0.47}$ & $0.59^{+0.40}_{-1.6}$ & $18.5\times10^3$ & 0.97\\
        \hline
        \multirow{4}{8em}{$-19.401\pm0.027$} & $137\pm4.5$ & $68.58\pm0.99$ & $-0.406\pm0.092$ & $0.68^{+0.36}_{-0.47}$ & $0.44^{+0.34}_{-1.5}$ & $2.8\times10^3$ & 1.08\\
        & $139.7\pm4.85$ & $68.44\pm0.99$ & $-0.413\pm0.092$ & $0.70^{+0.36}_{-0.48}$ & $0.49^{+0.39}_{-1.5}$ & $6.9\times10^3$ & 0.96\\
        & $147.05\pm0.3$ & $67.99\pm0.86$ & $-0.431\pm0.088$ & $0.76^{+0.36}_{-0.47}$ & $0.63^{+0.42}_{-1.6}$ & $3.5\times10^4$ & 0.62\\
        & $148\pm3.6$    & $68.09\pm0.96$ & $-0.427\pm0.090$ & $0.74^{+0.36}_{-0.47}$ & $0.59^{+0.42}_{-1.6}$ & $1.73\times10^4$ & 0.66\\
        \hline
        \multirow{4}{8em}{$-19.420\pm0.014$} & $137\pm4.5$ & $67.61\pm0.73$ & $-0.401\pm0.091$ & $0.68^{+0.36}_{-0.48}$ & $0.46^{+0.33}_{-1.5}$ & $1.3\times10^3$ & 0.28\\
        & $139.7\pm4.85$ & $67.58\pm0.73$ & $-0.409\pm0.092$ & $0.70^{+0.36}_{-0.48}$ & $0.51^{+0.36}_{-1.5}$ & $4.5\times10^3$ & 0.24\\
        & $147.05\pm0.3$ & $67.56\pm0.72$ & $-0.413\pm0.087$ & $0.71^{+0.35}_{-0.47}$ & $0.52^{+0.38}_{-1.5}$ & $4.3\times10^4$ & 0.22\\
        & $148\pm3.6$    & $67.54\pm0.73$ & $-0.424\pm0.090$ & $0.74^{+0.37}_{-0.47}$ & $0.61^{+0.42}_{-1.6}$ & $1.8\times10^4$ & 0.2\\
        \hline
    \end{tabular}
\end{table}

\section{Conclusions}
\label{sec:conclusion}
In this work, we have  investigated the effect of certain data-motivated priors (on the sound horizon $r_d$ as well as the peak absolute magnitude $M$) on the Hubble constant value and the potential impact on the Hubble tension. Our aim is not to address the Hubble tension but to show that degeneracies play a significant role \rthis{in some of the  low-redshift measurements of $H_0$}, and the usage of data-driven priors  are essential.

We emphasize that this work is heavily motivated by \citet{Chen_2024}. However, the aim of this work is to study the effect of degeneracy between $M$ and  $r_d$ on the \rthis{Hubble constant} values by applying data motivated Gaussian and uniform priors on the two parameters. In addition, \cite{Chen_2024} considered the Pantheon$+$ dataset only and $M$ is the only additional parameter other than $H_0$ and $\Omega_m$. In our analyses, we considered both $M$, $r_d$, as well as the  DESI BAO and CC datasets. We do wish to point out that the value of $\Omega_m$ had a very high value in \cite{Chen_2024}, while we get considerably lower values in comparison. Further, note that our $\Omega_m$ values are similar for all our different prior choices within $1\sigma$, irrespective of the values of $M$ or $r_d$. This was also noted in \cite{Chen_2024}, and occurs in our work because we consider only a particular redshift range (0.1 - 1.965).  

A summary of our key results can be found in Tables \ref{table2}, \ref{table3}, \ref{table4}, \ref{table5}, \ref{table6}, \ref{table7}, \ref{table8} and \ref{table9}.
We find that increasing the value of the sound horizon at the drag epoch does seem to reduce the \rthis{discrepancy} between the two values to somewhere around $1.2 \sigma$ ($\Lambda$CDM case), but the value of $M$ also decreases to about $-19.4$. This is also evident when we compare the Bayes' factors of other prior combinations with the prior combination consisting of SH0ES prior on $M$ and the CMB inferred prior on $r_d$, applied to a spatially-flat $\Lambda$CDM model.
This reaffirms  the fact that there is some degeneracy between $M$ and $r_d$, which needs to be further studied. Furthermore, we note that when applying uniform priors on either $r_d$ or $M$, the Hubble constant decreases for smaller values of $M$ and larger values of $r_d$. This is similar to the fact that Gaussian priors on both $r_d$ or $M$ decrease the value of $M$ and increase $r_d$. But note that $M$ plays a crucial role in decreasing $H_0$, as a higher $M$ and larger $r_d$ results in a tension of about $5\sigma$, while  smaller $M$ and smaller $r_d$ give a \rthis{discrepancy} of around $1.5\sigma$ (cf. Table~\ref{table8}). \rthis{The cosmography approach also give similar results as the $\Lambda$CDM case. However, the values of $H_0$ are slightly lower than the corresponding cases in the $\Lambda$CDM model approach and coupled with the large error bars drive the \rthis{discrepancy} to as low as $0.2\sigma$ (cf. Table~\ref{table9})}

Therefore, to summarize, we have reaffirmed the degeneracy present in the $M-r_d$ plane in light of the latest DESI results, and is in accord with some prior related works in literature~\cite{Dinda, denitsa, benitsy}.

\begin{acknowledgments}
SB would like to extend his gratitude to the University Grants Commission (UGC), Govt. of India for their continuous support through the Junior Research Fellowship, which has played a crucial role in the successful completion of our research. The authors also thank Bharat Ratra for useful comments on the manuscript. \rthis{We also acknowledge the anonymous referee for useful feedback and comments on the manuscript.}
\end{acknowledgments}

\bibliography{References}

\begin{thebibliography}{108}
\expandafter\ifx\csname natexlab\endcsname\relax\def\natexlab#1{#1}\fi
\expandafter\ifx\csname bibnamefont\endcsname\relax
  \def\bibnamefont#1{#1}\fi
\expandafter\ifx\csname bibfnamefont\endcsname\relax
  \def\bibfnamefont#1{#1}\fi
\expandafter\ifx\csname citenamefont\endcsname\relax
  \def\citenamefont#1{#1}\fi
\expandafter\ifx\csname url\endcsname\relax
  \def\url#1{\texttt{#1}}\fi
\expandafter\ifx\csname urlprefix\endcsname\relax\def\urlprefix{URL }\fi
\providecommand{\bibinfo}[2]{#2}
\providecommand{\eprint}[2][]{\url{#2}}

\bibitem[{\citenamefont{{Freedman} and {Madore}}(2010)}]{freedman}
\bibinfo{author}{\bibfnamefont{W.~L.} \bibnamefont{{Freedman}}} \bibnamefont{and} \bibinfo{author}{\bibfnamefont{B.~F.} \bibnamefont{{Madore}}}, \bibinfo{journal}{\araa} \textbf{\bibinfo{volume}{48}}, \bibinfo{pages}{673} (\bibinfo{year}{2010}), \eprint{1004.1856}.

\bibitem[{\citenamefont{{Hubble}}(1929)}]{hubble}
\bibinfo{author}{\bibfnamefont{E.}~\bibnamefont{{Hubble}}}, \bibinfo{journal}{Proceedings of the National Academy of Science} \textbf{\bibinfo{volume}{15}}, \bibinfo{pages}{168} (\bibinfo{year}{1929}).

\bibitem[{\citenamefont{{Chen} and {Ratra}}(2011)}]{ChanRatra}
\bibinfo{author}{\bibfnamefont{G.}~\bibnamefont{{Chen}}} \bibnamefont{and} \bibinfo{author}{\bibfnamefont{B.}~\bibnamefont{{Ratra}}}, \bibinfo{journal}{\pasp} \textbf{\bibinfo{volume}{123}}, \bibinfo{pages}{1127} (\bibinfo{year}{2011}), \eprint{1105.5206}.

\bibitem[{\citenamefont{{Bethapudi} and {Desai}}(2017)}]{Bethapudi}
\bibinfo{author}{\bibfnamefont{S.}~\bibnamefont{{Bethapudi}}} \bibnamefont{and} \bibinfo{author}{\bibfnamefont{S.}~\bibnamefont{{Desai}}}, \bibinfo{journal}{European Physical Journal Plus} \textbf{\bibinfo{volume}{132}}, \bibinfo{eid}{78} (\bibinfo{year}{2017}), \eprint{1701.01789}.

\bibitem[{\citenamefont{{Cervantes-Cota} et~al.}(2023)\citenamefont{{Cervantes-Cota}, {Galindo-Uribarri}, and {Smoot}}}]{Smoot23}
\bibinfo{author}{\bibfnamefont{J.~L.} \bibnamefont{{Cervantes-Cota}}}, \bibinfo{author}{\bibfnamefont{S.}~\bibnamefont{{Galindo-Uribarri}}}, \bibnamefont{and} \bibinfo{author}{\bibfnamefont{G.~F.} \bibnamefont{{Smoot}}}, \bibinfo{journal}{Universe} \textbf{\bibinfo{volume}{9}}, \bibinfo{eid}{501} (\bibinfo{year}{2023}), \eprint{2311.07552}.

\bibitem[{\citenamefont{{Freedman} et~al.}(2001)\citenamefont{{Freedman}, {Madore}, {Gibson}, {Ferrarese}, {Kelson}, {Sakai}, {Mould}, {Kennicutt}, {Ford}, {Graham} et~al.}}]{hst}
\bibinfo{author}{\bibfnamefont{W.~L.} \bibnamefont{{Freedman}}}, \bibinfo{author}{\bibfnamefont{B.~F.} \bibnamefont{{Madore}}}, \bibinfo{author}{\bibfnamefont{B.~K.} \bibnamefont{{Gibson}}}, \bibinfo{author}{\bibfnamefont{L.}~\bibnamefont{{Ferrarese}}}, \bibinfo{author}{\bibfnamefont{D.~D.} \bibnamefont{{Kelson}}}, \bibinfo{author}{\bibfnamefont{S.}~\bibnamefont{{Sakai}}}, \bibinfo{author}{\bibfnamefont{J.~R.} \bibnamefont{{Mould}}}, \bibinfo{author}{\bibfnamefont{R.~C.} \bibnamefont{{Kennicutt}}, \bibfnamefont{Jr.}}, \bibinfo{author}{\bibfnamefont{H.~C.} \bibnamefont{{Ford}}}, \bibinfo{author}{\bibfnamefont{J.~A.} \bibnamefont{{Graham}}}, \bibnamefont{et~al.}, \bibinfo{journal}{\apj} \textbf{\bibinfo{volume}{553}}, \bibinfo{pages}{47} (\bibinfo{year}{2001}), \eprint{astro-ph/0012376}.

\bibitem[{\citenamefont{{Planck Collaboration} et~al.}(2020)\citenamefont{{Planck Collaboration}, {Aghanim}, {Akrami}, {Ashdown}, {Aumont}, {Baccigalupi}, {Ballardini}, {Banday}, {Barreiro}, {Bartolo} et~al.}}]{Planck}
\bibinfo{author}{\bibnamefont{{Planck Collaboration}}}, \bibinfo{author}{\bibfnamefont{N.}~\bibnamefont{{Aghanim}}}, \bibinfo{author}{\bibfnamefont{Y.}~\bibnamefont{{Akrami}}}, \bibinfo{author}{\bibfnamefont{M.}~\bibnamefont{{Ashdown}}}, \bibinfo{author}{\bibfnamefont{J.}~\bibnamefont{{Aumont}}}, \bibinfo{author}{\bibfnamefont{C.}~\bibnamefont{{Baccigalupi}}}, \bibinfo{author}{\bibfnamefont{M.}~\bibnamefont{{Ballardini}}}, \bibinfo{author}{\bibfnamefont{A.~J.} \bibnamefont{{Banday}}}, \bibinfo{author}{\bibfnamefont{R.~B.} \bibnamefont{{Barreiro}}}, \bibinfo{author}{\bibfnamefont{N.}~\bibnamefont{{Bartolo}}}, \bibnamefont{et~al.}, \bibinfo{journal}{\aap} \textbf{\bibinfo{volume}{641}}, \bibinfo{eid}{A6} (\bibinfo{year}{2020}), \eprint{1807.06209}.

\bibitem[{\citenamefont{Riess et~al.}(2022)}]{shoes}
\bibinfo{author}{\bibfnamefont{A.~G.} \bibnamefont{Riess}} \bibnamefont{et~al.}, \bibinfo{journal}{Astrophys. J. Lett.} \textbf{\bibinfo{volume}{934}}, \bibinfo{pages}{L7} (\bibinfo{year}{2022}), \eprint{2112.04510}.

\bibitem[{\citenamefont{{Abdalla} et~al.}(2022{\natexlab{a}})\citenamefont{{Abdalla}, {Abell{\'a}n}, {Aboubrahim}, {Agnello}, {Akarsu}, {Akrami}, {Alestas}, {Aloni}, {Amendola}, {Anchordoqui} et~al.}}]{tensionreview}
\bibinfo{author}{\bibfnamefont{E.}~\bibnamefont{{Abdalla}}}, \bibinfo{author}{\bibfnamefont{G.~F.} \bibnamefont{{Abell{\'a}n}}}, \bibinfo{author}{\bibfnamefont{A.}~\bibnamefont{{Aboubrahim}}}, \bibinfo{author}{\bibfnamefont{A.}~\bibnamefont{{Agnello}}}, \bibinfo{author}{\bibfnamefont{{\"O}.}~\bibnamefont{{Akarsu}}}, \bibinfo{author}{\bibfnamefont{Y.}~\bibnamefont{{Akrami}}}, \bibinfo{author}{\bibfnamefont{G.}~\bibnamefont{{Alestas}}}, \bibinfo{author}{\bibfnamefont{D.}~\bibnamefont{{Aloni}}}, \bibinfo{author}{\bibfnamefont{L.}~\bibnamefont{{Amendola}}}, \bibinfo{author}{\bibfnamefont{L.~A.} \bibnamefont{{Anchordoqui}}}, \bibnamefont{et~al.}, \bibinfo{journal}{Journal of High Energy Astrophysics} \textbf{\bibinfo{volume}{34}}, \bibinfo{pages}{49} (\bibinfo{year}{2022}{\natexlab{a}}), \eprint{2203.06142}.

\bibitem[{\citenamefont{{Verde} et~al.}(2019)\citenamefont{{Verde}, {Treu}, and {Riess}}}]{Verde}
\bibinfo{author}{\bibfnamefont{L.}~\bibnamefont{{Verde}}}, \bibinfo{author}{\bibfnamefont{T.}~\bibnamefont{{Treu}}}, \bibnamefont{and} \bibinfo{author}{\bibfnamefont{A.~G.} \bibnamefont{{Riess}}}, \bibinfo{journal}{Nature Astronomy} \textbf{\bibinfo{volume}{3}}, \bibinfo{pages}{891} (\bibinfo{year}{2019}), \eprint{1907.10625}.

\bibitem[{\citenamefont{{Di Valentino} et~al.}(2021)\citenamefont{{Di Valentino}, {Mena}, {Pan}, {Visinelli}, {Yang}, {Melchiorri}, {Mota}, {Riess}, and {Silk}}}]{DiValentino22}
\bibinfo{author}{\bibfnamefont{E.}~\bibnamefont{{Di Valentino}}}, \bibinfo{author}{\bibfnamefont{O.}~\bibnamefont{{Mena}}}, \bibinfo{author}{\bibfnamefont{S.}~\bibnamefont{{Pan}}}, \bibinfo{author}{\bibfnamefont{L.}~\bibnamefont{{Visinelli}}}, \bibinfo{author}{\bibfnamefont{W.}~\bibnamefont{{Yang}}}, \bibinfo{author}{\bibfnamefont{A.}~\bibnamefont{{Melchiorri}}}, \bibinfo{author}{\bibfnamefont{D.~F.} \bibnamefont{{Mota}}}, \bibinfo{author}{\bibfnamefont{A.~G.} \bibnamefont{{Riess}}}, \bibnamefont{and} \bibinfo{author}{\bibfnamefont{J.}~\bibnamefont{{Silk}}}, \bibinfo{journal}{Classical and Quantum Gravity} \textbf{\bibinfo{volume}{38}}, \bibinfo{eid}{153001} (\bibinfo{year}{2021}), \eprint{2103.01183}.

\bibitem[{\citenamefont{{Knox} and {Millea}}(2020)}]{hubblehunter}
\bibinfo{author}{\bibfnamefont{L.}~\bibnamefont{{Knox}}} \bibnamefont{and} \bibinfo{author}{\bibfnamefont{M.}~\bibnamefont{{Millea}}}, \bibinfo{journal}{\prd} \textbf{\bibinfo{volume}{101}}, \bibinfo{eid}{043533} (\bibinfo{year}{2020}), \eprint{1908.03663}.

\bibitem[{\citenamefont{{Poulin} et~al.}(2023)\citenamefont{{Poulin}, {Smith}, and {Karwal}}}]{poulin}
\bibinfo{author}{\bibfnamefont{V.}~\bibnamefont{{Poulin}}}, \bibinfo{author}{\bibfnamefont{T.~L.} \bibnamefont{{Smith}}}, \bibnamefont{and} \bibinfo{author}{\bibfnamefont{T.}~\bibnamefont{{Karwal}}}, \bibinfo{journal}{Physics of the Dark Universe} \textbf{\bibinfo{volume}{42}}, \bibinfo{eid}{101348} (\bibinfo{year}{2023}), \eprint{2302.09032}.

\bibitem[{\citenamefont{{Karwal} et~al.}(2022)\citenamefont{{Karwal}, {Raveri}, {Jain}, {Khoury}, and {Trodden}}}]{karwal}
\bibinfo{author}{\bibfnamefont{T.}~\bibnamefont{{Karwal}}}, \bibinfo{author}{\bibfnamefont{M.}~\bibnamefont{{Raveri}}}, \bibinfo{author}{\bibfnamefont{B.}~\bibnamefont{{Jain}}}, \bibinfo{author}{\bibfnamefont{J.}~\bibnamefont{{Khoury}}}, \bibnamefont{and} \bibinfo{author}{\bibfnamefont{M.}~\bibnamefont{{Trodden}}}, \bibinfo{journal}{\prd} \textbf{\bibinfo{volume}{105}}, \bibinfo{eid}{063535} (\bibinfo{year}{2022}), \eprint{2106.13290}.

\bibitem[{\citenamefont{{Poulin} et~al.}(2019)\citenamefont{{Poulin}, {Smith}, {Karwal}, and {Kamionkowski}}}]{poulin_solution}
\bibinfo{author}{\bibfnamefont{V.}~\bibnamefont{{Poulin}}}, \bibinfo{author}{\bibfnamefont{T.~L.} \bibnamefont{{Smith}}}, \bibinfo{author}{\bibfnamefont{T.}~\bibnamefont{{Karwal}}}, \bibnamefont{and} \bibinfo{author}{\bibfnamefont{M.}~\bibnamefont{{Kamionkowski}}}, \bibinfo{journal}{\prl} \textbf{\bibinfo{volume}{122}}, \bibinfo{eid}{221301} (\bibinfo{year}{2019}), \eprint{1811.04083}.

\bibitem[{\citenamefont{{Berghaus} and {Karwal}}(2020)}]{berghaus}
\bibinfo{author}{\bibfnamefont{K.~V.} \bibnamefont{{Berghaus}}} \bibnamefont{and} \bibinfo{author}{\bibfnamefont{T.}~\bibnamefont{{Karwal}}}, \bibinfo{journal}{\prd} \textbf{\bibinfo{volume}{101}}, \bibinfo{eid}{083537} (\bibinfo{year}{2020}), \eprint{1911.06281}.

\bibitem[{\citenamefont{{Kamionkowski} and {Riess}}(2023)}]{KamionkowskiRiess}
\bibinfo{author}{\bibfnamefont{M.}~\bibnamefont{{Kamionkowski}}} \bibnamefont{and} \bibinfo{author}{\bibfnamefont{A.~G.} \bibnamefont{{Riess}}}, \bibinfo{journal}{Annual Review of Nuclear and Particle Science} \textbf{\bibinfo{volume}{73}}, \bibinfo{pages}{153} (\bibinfo{year}{2023}), \eprint{2211.04492}.

\bibitem[{\citenamefont{{Agrawal} et~al.}(2023)\citenamefont{{Agrawal}, {Cyr-Racine}, {Pinner}, and {Randall}}}]{agrawal}
\bibinfo{author}{\bibfnamefont{P.}~\bibnamefont{{Agrawal}}}, \bibinfo{author}{\bibfnamefont{F.-Y.} \bibnamefont{{Cyr-Racine}}}, \bibinfo{author}{\bibfnamefont{D.}~\bibnamefont{{Pinner}}}, \bibnamefont{and} \bibinfo{author}{\bibfnamefont{L.}~\bibnamefont{{Randall}}}, \bibinfo{journal}{Physics of the Dark Universe} \textbf{\bibinfo{volume}{42}}, \bibinfo{eid}{101347} (\bibinfo{year}{2023}), \eprint{1904.01016}.

\bibitem[{\citenamefont{Vagnozzi}(2023)}]{Vagnozzi}
\bibinfo{author}{\bibfnamefont{S.}~\bibnamefont{Vagnozzi}}, \bibinfo{journal}{Universe} \textbf{\bibinfo{volume}{9}}, \bibinfo{pages}{393} (\bibinfo{year}{2023}), \eprint{2308.16628}.

\bibitem[{\citenamefont{{Montani} et~al.}(2025{\natexlab{a}})\citenamefont{{Montani}, {Carlevaro}, {Escamilla}, and {Di Valentino}}}]{Montani}
\bibinfo{author}{\bibfnamefont{G.}~\bibnamefont{{Montani}}}, \bibinfo{author}{\bibfnamefont{N.}~\bibnamefont{{Carlevaro}}}, \bibinfo{author}{\bibfnamefont{L.~A.} \bibnamefont{{Escamilla}}}, \bibnamefont{and} \bibinfo{author}{\bibfnamefont{E.}~\bibnamefont{{Di Valentino}}}, \bibinfo{journal}{Physics of the Dark Universe} \textbf{\bibinfo{volume}{48}}, \bibinfo{eid}{101848} (\bibinfo{year}{2025}{\natexlab{a}}), \eprint{2404.15977}.

\bibitem[{\citenamefont{{Raveri}}(2020)}]{raveri}
\bibinfo{author}{\bibfnamefont{M.}~\bibnamefont{{Raveri}}}, \bibinfo{journal}{\prd} \textbf{\bibinfo{volume}{101}}, \bibinfo{eid}{083524} (\bibinfo{year}{2020}), \eprint{1902.01366}.

\bibitem[{\citenamefont{{Keeley} et~al.}(2019)\citenamefont{{Keeley}, {Joudaki}, {Kaplinghat}, and {Kirkby}}}]{keeley}
\bibinfo{author}{\bibfnamefont{R.~E.} \bibnamefont{{Keeley}}}, \bibinfo{author}{\bibfnamefont{S.}~\bibnamefont{{Joudaki}}}, \bibinfo{author}{\bibfnamefont{M.}~\bibnamefont{{Kaplinghat}}}, \bibnamefont{and} \bibinfo{author}{\bibfnamefont{D.}~\bibnamefont{{Kirkby}}}, \bibinfo{journal}{\jcap} \textbf{\bibinfo{volume}{2019}}, \bibinfo{eid}{035} (\bibinfo{year}{2019}), \eprint{1905.10198}.

\bibitem[{\citenamefont{{Colg{\'a}in} and {Sheikh-Jabbari}}(2024)}]{Colgain24}
\bibinfo{author}{\bibfnamefont{E.~{\'O}.} \bibnamefont{{Colg{\'a}in}}} \bibnamefont{and} \bibinfo{author}{\bibfnamefont{M.~M.} \bibnamefont{{Sheikh-Jabbari}}}, \bibinfo{journal}{arXiv e-prints} \bibinfo{eid}{arXiv:2412.12905} (\bibinfo{year}{2024}), \eprint{2412.12905}.

\bibitem[{\citenamefont{{Dainotti} et~al.}(2021)\citenamefont{{Dainotti}, {De Simone}, {Schiavone}, {Montani}, {Rinaldi}, and {Lambiase}}}]{dainotti_2021}
\bibinfo{author}{\bibfnamefont{M.~G.} \bibnamefont{{Dainotti}}}, \bibinfo{author}{\bibfnamefont{B.}~\bibnamefont{{De Simone}}}, \bibinfo{author}{\bibfnamefont{T.}~\bibnamefont{{Schiavone}}}, \bibinfo{author}{\bibfnamefont{G.}~\bibnamefont{{Montani}}}, \bibinfo{author}{\bibfnamefont{E.}~\bibnamefont{{Rinaldi}}}, \bibnamefont{and} \bibinfo{author}{\bibfnamefont{G.}~\bibnamefont{{Lambiase}}}, \bibinfo{journal}{\apj} \textbf{\bibinfo{volume}{912}}, \bibinfo{eid}{150} (\bibinfo{year}{2021}), \eprint{2103.02117}.

\bibitem[{\citenamefont{{Montani} et~al.}(2024)\citenamefont{{Montani}, {Carlevaro}, and {Dainotti}}}]{Montani24}
\bibinfo{author}{\bibfnamefont{G.}~\bibnamefont{{Montani}}}, \bibinfo{author}{\bibfnamefont{N.}~\bibnamefont{{Carlevaro}}}, \bibnamefont{and} \bibinfo{author}{\bibfnamefont{M.~G.} \bibnamefont{{Dainotti}}}, \bibinfo{journal}{Physics of the Dark Universe} \textbf{\bibinfo{volume}{44}}, \bibinfo{eid}{101486} (\bibinfo{year}{2024}), \eprint{2311.04822}.

\bibitem[{\citenamefont{{Montani} et~al.}(2025{\natexlab{b}})\citenamefont{{Montani}, {Carlevaro}, and {Dainotti}}}]{montani_2025}
\bibinfo{author}{\bibfnamefont{G.}~\bibnamefont{{Montani}}}, \bibinfo{author}{\bibfnamefont{N.}~\bibnamefont{{Carlevaro}}}, \bibnamefont{and} \bibinfo{author}{\bibfnamefont{M.~G.} \bibnamefont{{Dainotti}}}, \bibinfo{journal}{Physics of the Dark Universe} \textbf{\bibinfo{volume}{48}}, \bibinfo{eid}{101847} (\bibinfo{year}{2025}{\natexlab{b}}), \eprint{2411.07060}.

\bibitem[{\citenamefont{{Perivolaropoulos} and {Skara}}(2022{\natexlab{a}})}]{Periv}
\bibinfo{author}{\bibfnamefont{L.}~\bibnamefont{{Perivolaropoulos}}} \bibnamefont{and} \bibinfo{author}{\bibfnamefont{F.}~\bibnamefont{{Skara}}}, \bibinfo{journal}{\nar} \textbf{\bibinfo{volume}{95}}, \bibinfo{eid}{101659} (\bibinfo{year}{2022}{\natexlab{a}}), \eprint{2105.05208}.

\bibitem[{\citenamefont{{Abdalla} et~al.}(2022{\natexlab{b}})\citenamefont{{Abdalla}, {Abell{\'a}n}, {Aboubrahim}, {Agnello}, {Akarsu}, {Akrami}, {Alestas}, {Aloni}, {Amendola}, {Anchordoqui} et~al.}}]{Abdalla22}
\bibinfo{author}{\bibfnamefont{E.}~\bibnamefont{{Abdalla}}}, \bibinfo{author}{\bibfnamefont{G.~F.} \bibnamefont{{Abell{\'a}n}}}, \bibinfo{author}{\bibfnamefont{A.}~\bibnamefont{{Aboubrahim}}}, \bibinfo{author}{\bibfnamefont{A.}~\bibnamefont{{Agnello}}}, \bibinfo{author}{\bibfnamefont{{\"O}.}~\bibnamefont{{Akarsu}}}, \bibinfo{author}{\bibfnamefont{Y.}~\bibnamefont{{Akrami}}}, \bibinfo{author}{\bibfnamefont{G.}~\bibnamefont{{Alestas}}}, \bibinfo{author}{\bibfnamefont{D.}~\bibnamefont{{Aloni}}}, \bibinfo{author}{\bibfnamefont{L.}~\bibnamefont{{Amendola}}}, \bibinfo{author}{\bibfnamefont{L.~A.} \bibnamefont{{Anchordoqui}}}, \bibnamefont{et~al.}, \bibinfo{journal}{Journal of High Energy Astrophysics} \textbf{\bibinfo{volume}{34}}, \bibinfo{pages}{49} (\bibinfo{year}{2022}{\natexlab{b}}), \eprint{2203.06142}.

\bibitem[{\citenamefont{{Peebles}}(2022)}]{Peebles22}
\bibinfo{author}{\bibfnamefont{P.~J.~E.} \bibnamefont{{Peebles}}}, \bibinfo{journal}{Annals of Physics} \textbf{\bibinfo{volume}{447}}, \bibinfo{eid}{169159} (\bibinfo{year}{2022}), \eprint{2208.05018}.

\bibitem[{\citenamefont{{Banik} and {Zhao}}(2022)}]{Banik}
\bibinfo{author}{\bibfnamefont{I.}~\bibnamefont{{Banik}}} \bibnamefont{and} \bibinfo{author}{\bibfnamefont{H.}~\bibnamefont{{Zhao}}}, \bibinfo{journal}{Symmetry} \textbf{\bibinfo{volume}{14}}, \bibinfo{pages}{1331} (\bibinfo{year}{2022}), \eprint{2110.06936}.

\bibitem[{\citenamefont{{Eisenstein} and {Hu}}(1998)}]{Eisenstein}
\bibinfo{author}{\bibfnamefont{D.~J.} \bibnamefont{{Eisenstein}}} \bibnamefont{and} \bibinfo{author}{\bibfnamefont{W.}~\bibnamefont{{Hu}}}, \bibinfo{journal}{\apj} \textbf{\bibinfo{volume}{496}}, \bibinfo{pages}{605} (\bibinfo{year}{1998}), \eprint{astro-ph/9709112}.

\bibitem[{\citenamefont{{Sunyaev} and {Zeldovich}}(1972)}]{sunyaev}
\bibinfo{author}{\bibfnamefont{R.~A.} \bibnamefont{{Sunyaev}}} \bibnamefont{and} \bibinfo{author}{\bibfnamefont{Y.~B.} \bibnamefont{{Zeldovich}}}, \bibinfo{journal}{Comments on Astrophysics and Space Physics} \textbf{\bibinfo{volume}{4}}, \bibinfo{pages}{173} (\bibinfo{year}{1972}).

\bibitem[{\citenamefont{{Peebles} and {Yu}}(1970)}]{Peebles}
\bibinfo{author}{\bibfnamefont{P.~J.~E.} \bibnamefont{{Peebles}}} \bibnamefont{and} \bibinfo{author}{\bibfnamefont{J.~T.} \bibnamefont{{Yu}}}, \bibinfo{journal}{\apj} \textbf{\bibinfo{volume}{162}}, \bibinfo{pages}{815} (\bibinfo{year}{1970}).

\bibitem[{\citenamefont{{Eisenstein} et~al.}(2007)\citenamefont{{Eisenstein}, {Seo}, {Sirko}, and {Spergel}}}]{eisensteinspergel}
\bibinfo{author}{\bibfnamefont{D.~J.} \bibnamefont{{Eisenstein}}}, \bibinfo{author}{\bibfnamefont{H.-J.} \bibnamefont{{Seo}}}, \bibinfo{author}{\bibfnamefont{E.}~\bibnamefont{{Sirko}}}, \bibnamefont{and} \bibinfo{author}{\bibfnamefont{D.~N.} \bibnamefont{{Spergel}}}, \bibinfo{journal}{\apj} \textbf{\bibinfo{volume}{664}}, \bibinfo{pages}{675} (\bibinfo{year}{2007}), \eprint{astro-ph/0604362}.

\bibitem[{\citenamefont{{Sutherland}}(2012)}]{sutherland}
\bibinfo{author}{\bibfnamefont{W.}~\bibnamefont{{Sutherland}}}, \bibinfo{journal}{\mnras} \textbf{\bibinfo{volume}{426}}, \bibinfo{pages}{1280} (\bibinfo{year}{2012}), \eprint{1205.0715}.

\bibitem[{\citenamefont{{Riess} et~al.}(1998)\citenamefont{{Riess}, {Filippenko}, {Challis}, {Clocchiatti}, {Diercks}, {Garnavich}, {Gilliland}, {Hogan}, {Jha}, {Kirshner} et~al.}}]{riess1998}
\bibinfo{author}{\bibfnamefont{A.~G.} \bibnamefont{{Riess}}}, \bibinfo{author}{\bibfnamefont{A.~V.} \bibnamefont{{Filippenko}}}, \bibinfo{author}{\bibfnamefont{P.}~\bibnamefont{{Challis}}}, \bibinfo{author}{\bibfnamefont{A.}~\bibnamefont{{Clocchiatti}}}, \bibinfo{author}{\bibfnamefont{A.}~\bibnamefont{{Diercks}}}, \bibinfo{author}{\bibfnamefont{P.~M.} \bibnamefont{{Garnavich}}}, \bibinfo{author}{\bibfnamefont{R.~L.} \bibnamefont{{Gilliland}}}, \bibinfo{author}{\bibfnamefont{C.~J.} \bibnamefont{{Hogan}}}, \bibinfo{author}{\bibfnamefont{S.}~\bibnamefont{{Jha}}}, \bibinfo{author}{\bibfnamefont{R.~P.} \bibnamefont{{Kirshner}}}, \bibnamefont{et~al.}, \bibinfo{journal}{\aj} \textbf{\bibinfo{volume}{116}}, \bibinfo{pages}{1009} (\bibinfo{year}{1998}), \eprint{astro-ph/9805201}.

\bibitem[{\citenamefont{{Perlmutter} et~al.}(1999)\citenamefont{{Perlmutter}, {Aldering}, {Goldhaber}, {Knop}, {Nugent}, {Castro}, {Deustua}, {Fabbro}, {Goobar}, {Groom} et~al.}}]{Perlmutter}
\bibinfo{author}{\bibfnamefont{S.}~\bibnamefont{{Perlmutter}}}, \bibinfo{author}{\bibfnamefont{G.}~\bibnamefont{{Aldering}}}, \bibinfo{author}{\bibfnamefont{G.}~\bibnamefont{{Goldhaber}}}, \bibinfo{author}{\bibfnamefont{R.~A.} \bibnamefont{{Knop}}}, \bibinfo{author}{\bibfnamefont{P.}~\bibnamefont{{Nugent}}}, \bibinfo{author}{\bibfnamefont{P.~G.} \bibnamefont{{Castro}}}, \bibinfo{author}{\bibfnamefont{S.}~\bibnamefont{{Deustua}}}, \bibinfo{author}{\bibfnamefont{S.}~\bibnamefont{{Fabbro}}}, \bibinfo{author}{\bibfnamefont{A.}~\bibnamefont{{Goobar}}}, \bibinfo{author}{\bibfnamefont{D.~E.} \bibnamefont{{Groom}}}, \bibnamefont{et~al.}, \bibinfo{journal}{\apj} \textbf{\bibinfo{volume}{517}}, \bibinfo{pages}{565} (\bibinfo{year}{1999}), \eprint{astro-ph/9812133}.

\bibitem[{\citenamefont{{Huterer} and {Shafer}}(2018)}]{Huterer}
\bibinfo{author}{\bibfnamefont{D.}~\bibnamefont{{Huterer}}} \bibnamefont{and} \bibinfo{author}{\bibfnamefont{D.~L.} \bibnamefont{{Shafer}}}, \bibinfo{journal}{Reports on Progress in Physics} \textbf{\bibinfo{volume}{81}}, \bibinfo{eid}{016901} (\bibinfo{year}{2018}), \eprint{1709.01091}.

\bibitem[{\citenamefont{{Weinberg} et~al.}(2013)\citenamefont{{Weinberg}, {Mortonson}, {Eisenstein}, {Hirata}, {Riess}, and {Rozo}}}]{weinberg}
\bibinfo{author}{\bibfnamefont{D.~H.} \bibnamefont{{Weinberg}}}, \bibinfo{author}{\bibfnamefont{M.~J.} \bibnamefont{{Mortonson}}}, \bibinfo{author}{\bibfnamefont{D.~J.} \bibnamefont{{Eisenstein}}}, \bibinfo{author}{\bibfnamefont{C.}~\bibnamefont{{Hirata}}}, \bibinfo{author}{\bibfnamefont{A.~G.} \bibnamefont{{Riess}}}, \bibnamefont{and} \bibinfo{author}{\bibfnamefont{E.}~\bibnamefont{{Rozo}}}, \bibinfo{journal}{\physrep} \textbf{\bibinfo{volume}{530}}, \bibinfo{pages}{87} (\bibinfo{year}{2013}), \eprint{1201.2434}.

\bibitem[{\citenamefont{{Efstathiou}}(2021)}]{Efstathiou}
\bibinfo{author}{\bibfnamefont{G.}~\bibnamefont{{Efstathiou}}}, \bibinfo{journal}{\mnras} \textbf{\bibinfo{volume}{505}}, \bibinfo{pages}{3866} (\bibinfo{year}{2021}), \eprint{2103.08723}.

\bibitem[{\citenamefont{{Camarena} and {Marra}}(2021)}]{Camarena}
\bibinfo{author}{\bibfnamefont{D.}~\bibnamefont{{Camarena}}} \bibnamefont{and} \bibinfo{author}{\bibfnamefont{V.}~\bibnamefont{{Marra}}}, \bibinfo{journal}{\mnras} \textbf{\bibinfo{volume}{504}}, \bibinfo{pages}{5164} (\bibinfo{year}{2021}), \eprint{2101.08641}.

\bibitem[{\citenamefont{{Perivolaropoulos} and {Skara}}(2022{\natexlab{b}})}]{Perivolaropoulos1}
\bibinfo{author}{\bibfnamefont{L.}~\bibnamefont{{Perivolaropoulos}}} \bibnamefont{and} \bibinfo{author}{\bibfnamefont{F.}~\bibnamefont{{Skara}}}, \bibinfo{journal}{Universe} \textbf{\bibinfo{volume}{8}}, \bibinfo{eid}{502} (\bibinfo{year}{2022}{\natexlab{b}}), \eprint{2208.11169}.

\bibitem[{\citenamefont{{Perivolaropoulos} and {Skara}}(2023)}]{Perivolaropoulos2}
\bibinfo{author}{\bibfnamefont{L.}~\bibnamefont{{Perivolaropoulos}}} \bibnamefont{and} \bibinfo{author}{\bibfnamefont{F.}~\bibnamefont{{Skara}}}, \bibinfo{journal}{\mnras} \textbf{\bibinfo{volume}{520}}, \bibinfo{pages}{5110} (\bibinfo{year}{2023}), \eprint{2301.01024}.

\bibitem[{\citenamefont{{Ashall} et~al.}(2016)\citenamefont{{Ashall}, {Mazzali}, {Sasdelli}, and {Prentice}}}]{Ashall}
\bibinfo{author}{\bibfnamefont{C.}~\bibnamefont{{Ashall}}}, \bibinfo{author}{\bibfnamefont{P.}~\bibnamefont{{Mazzali}}}, \bibinfo{author}{\bibfnamefont{M.}~\bibnamefont{{Sasdelli}}}, \bibnamefont{and} \bibinfo{author}{\bibfnamefont{S.~J.} \bibnamefont{{Prentice}}}, \bibinfo{journal}{\mnras} \textbf{\bibinfo{volume}{460}}, \bibinfo{pages}{3529} (\bibinfo{year}{2016}), \eprint{1605.05507}.

\bibitem[{\citenamefont{{Evslin}}(2016)}]{Evslin}
\bibinfo{author}{\bibfnamefont{J.}~\bibnamefont{{Evslin}}}, \bibinfo{journal}{Physics of the Dark Universe} \textbf{\bibinfo{volume}{14}}, \bibinfo{pages}{57} (\bibinfo{year}{2016}), \eprint{1605.00486}.

\bibitem[{\citenamefont{{Alestas} et~al.}(2021)\citenamefont{{Alestas}, {Kazantzidis}, and {Perivolaropoulos}}}]{Alestas}
\bibinfo{author}{\bibfnamefont{G.}~\bibnamefont{{Alestas}}}, \bibinfo{author}{\bibfnamefont{L.}~\bibnamefont{{Kazantzidis}}}, \bibnamefont{and} \bibinfo{author}{\bibfnamefont{L.}~\bibnamefont{{Perivolaropoulos}}}, \bibinfo{journal}{\prd} \textbf{\bibinfo{volume}{103}}, \bibinfo{eid}{083517} (\bibinfo{year}{2021}), \eprint{2012.13932}.

\bibitem[{\citenamefont{{Staicova}}(2024)}]{denitsa}
\bibinfo{author}{\bibfnamefont{D.}~\bibnamefont{{Staicova}}}, \bibinfo{journal}{arXiv e-prints} \bibinfo{eid}{arXiv:2404.07182} (\bibinfo{year}{2024}), \eprint{2404.07182}.

\bibitem[{\citenamefont{{Mukherjee} et~al.}(2024)\citenamefont{{Mukherjee}, {Dialektopoulos}, {Said}, and {Mifsud}}}]{purba}
\bibinfo{author}{\bibfnamefont{P.}~\bibnamefont{{Mukherjee}}}, \bibinfo{author}{\bibfnamefont{K.~F.} \bibnamefont{{Dialektopoulos}}}, \bibinfo{author}{\bibfnamefont{J.~L.} \bibnamefont{{Said}}}, \bibnamefont{and} \bibinfo{author}{\bibfnamefont{J.}~\bibnamefont{{Mifsud}}}, \bibinfo{journal}{\jcap} \textbf{\bibinfo{volume}{2024}}, \bibinfo{eid}{060} (\bibinfo{year}{2024}), \eprint{2402.10502}.

\bibitem[{\citenamefont{{DESI Collaboration} et~al.}(2024)\citenamefont{{DESI Collaboration}, {Adame}, {Aguilar}, {Ahlen}, {Alam}, {Alexander}, {Alvarez}, {Alves}, {Anand}, {Andrade} et~al.}}]{DESI}
\bibinfo{author}{\bibnamefont{{DESI Collaboration}}}, \bibinfo{author}{\bibfnamefont{A.~G.} \bibnamefont{{Adame}}}, \bibinfo{author}{\bibfnamefont{J.}~\bibnamefont{{Aguilar}}}, \bibinfo{author}{\bibfnamefont{S.}~\bibnamefont{{Ahlen}}}, \bibinfo{author}{\bibfnamefont{S.}~\bibnamefont{{Alam}}}, \bibinfo{author}{\bibfnamefont{D.~M.} \bibnamefont{{Alexander}}}, \bibinfo{author}{\bibfnamefont{M.}~\bibnamefont{{Alvarez}}}, \bibinfo{author}{\bibfnamefont{O.}~\bibnamefont{{Alves}}}, \bibinfo{author}{\bibfnamefont{A.}~\bibnamefont{{Anand}}}, \bibinfo{author}{\bibfnamefont{U.}~\bibnamefont{{Andrade}}}, \bibnamefont{et~al.}, \bibinfo{journal}{arXiv e-prints} \bibinfo{eid}{arXiv:2404.03002} (\bibinfo{year}{2024}), \eprint{2404.03002}.

\bibitem[{\citenamefont{{Particle Data Group} et~al.}(2020)\citenamefont{{Particle Data Group}, {Zyla}, {Barnett}, {Beringer}, {Dahl}, {Dwyer}, {Groom}, {Lin}, {Lugovsky}, {Pianori} et~al.}}]{PDG}
\bibinfo{author}{\bibnamefont{{Particle Data Group}}}, \bibinfo{author}{\bibfnamefont{P.~A.} \bibnamefont{{Zyla}}}, \bibinfo{author}{\bibfnamefont{R.~M.} \bibnamefont{{Barnett}}}, \bibinfo{author}{\bibfnamefont{J.}~\bibnamefont{{Beringer}}}, \bibinfo{author}{\bibfnamefont{O.}~\bibnamefont{{Dahl}}}, \bibinfo{author}{\bibfnamefont{D.~A.} \bibnamefont{{Dwyer}}}, \bibinfo{author}{\bibfnamefont{D.~E.} \bibnamefont{{Groom}}}, \bibinfo{author}{\bibfnamefont{C.~J.} \bibnamefont{{Lin}}}, \bibinfo{author}{\bibfnamefont{K.~S.} \bibnamefont{{Lugovsky}}}, \bibinfo{author}{\bibfnamefont{E.}~\bibnamefont{{Pianori}}}, \bibnamefont{et~al.}, \bibinfo{journal}{Progress of Theoretical and Experimental Physics} \textbf{\bibinfo{volume}{2020}}, \bibinfo{eid}{083C01} (\bibinfo{year}{2020}).

\bibitem[{\citenamefont{{Aubourg} et~al.}(2015)\citenamefont{{Aubourg}, {Bailey}, {Bautista}, {Beutler}, {Bhardwaj}, {Bizyaev}, {Blanton}, {Blomqvist}, {Bolton}, {Bovy} et~al.}}]{augbourg}
\bibinfo{author}{\bibfnamefont{{\'E}.}~\bibnamefont{{Aubourg}}}, \bibinfo{author}{\bibfnamefont{S.}~\bibnamefont{{Bailey}}}, \bibinfo{author}{\bibfnamefont{J.~E.} \bibnamefont{{Bautista}}}, \bibinfo{author}{\bibfnamefont{F.}~\bibnamefont{{Beutler}}}, \bibinfo{author}{\bibfnamefont{V.}~\bibnamefont{{Bhardwaj}}}, \bibinfo{author}{\bibfnamefont{D.}~\bibnamefont{{Bizyaev}}}, \bibinfo{author}{\bibfnamefont{M.}~\bibnamefont{{Blanton}}}, \bibinfo{author}{\bibfnamefont{M.}~\bibnamefont{{Blomqvist}}}, \bibinfo{author}{\bibfnamefont{A.~S.} \bibnamefont{{Bolton}}}, \bibinfo{author}{\bibfnamefont{J.}~\bibnamefont{{Bovy}}}, \bibnamefont{et~al.}, \bibinfo{journal}{\prd} \textbf{\bibinfo{volume}{92}}, \bibinfo{eid}{123516} (\bibinfo{year}{2015}), \eprint{1411.1074}.

\bibitem[{\citenamefont{{Seo} and {Eisenstein}}(2003)}]{seo}
\bibinfo{author}{\bibfnamefont{H.-J.} \bibnamefont{{Seo}}} \bibnamefont{and} \bibinfo{author}{\bibfnamefont{D.~J.} \bibnamefont{{Eisenstein}}}, \bibinfo{journal}{\apj} \textbf{\bibinfo{volume}{598}}, \bibinfo{pages}{720} (\bibinfo{year}{2003}), \eprint{astro-ph/0307460}.

\bibitem[{\citenamefont{{Bora} and {Desai}}(2021)}]{BoraCDDR}
\bibinfo{author}{\bibfnamefont{K.}~\bibnamefont{{Bora}}} \bibnamefont{and} \bibinfo{author}{\bibfnamefont{S.}~\bibnamefont{{Desai}}}, \bibinfo{journal}{\jcap} \textbf{\bibinfo{volume}{2021}}, \bibinfo{eid}{052} (\bibinfo{year}{2021}), \eprint{2104.00974}.

\bibitem[{\citenamefont{{Eisenstein} et~al.}(2005)\citenamefont{{Eisenstein}, {Zehavi}, {Hogg}, {Scoccimarro}, {Blanton}, {Nichol}, {Scranton}, {Seo}, {Tegmark}, {Zheng} et~al.}}]{eisensteindv}
\bibinfo{author}{\bibfnamefont{D.~J.} \bibnamefont{{Eisenstein}}}, \bibinfo{author}{\bibfnamefont{I.}~\bibnamefont{{Zehavi}}}, \bibinfo{author}{\bibfnamefont{D.~W.} \bibnamefont{{Hogg}}}, \bibinfo{author}{\bibfnamefont{R.}~\bibnamefont{{Scoccimarro}}}, \bibinfo{author}{\bibfnamefont{M.~R.} \bibnamefont{{Blanton}}}, \bibinfo{author}{\bibfnamefont{R.~C.} \bibnamefont{{Nichol}}}, \bibinfo{author}{\bibfnamefont{R.}~\bibnamefont{{Scranton}}}, \bibinfo{author}{\bibfnamefont{H.-J.} \bibnamefont{{Seo}}}, \bibinfo{author}{\bibfnamefont{M.}~\bibnamefont{{Tegmark}}}, \bibinfo{author}{\bibfnamefont{Z.}~\bibnamefont{{Zheng}}}, \bibnamefont{et~al.}, \bibinfo{journal}{\apj} \textbf{\bibinfo{volume}{633}}, \bibinfo{pages}{560} (\bibinfo{year}{2005}), \eprint{astro-ph/0501171}.

\bibitem[{\citenamefont{{Petreca} et~al.}(2024)\citenamefont{{Petreca}, {Benetti}, and {Capozziello}}}]{petreca_2024}
\bibinfo{author}{\bibfnamefont{A.~T.} \bibnamefont{{Petreca}}}, \bibinfo{author}{\bibfnamefont{M.}~\bibnamefont{{Benetti}}}, \bibnamefont{and} \bibinfo{author}{\bibfnamefont{S.}~\bibnamefont{{Capozziello}}}, \bibinfo{journal}{Physics of the Dark Universe} \textbf{\bibinfo{volume}{44}}, \bibinfo{eid}{101453} (\bibinfo{year}{2024}), \eprint{2309.15711}.

\bibitem[{\citenamefont{{Aviles} et~al.}(2014)\citenamefont{{Aviles}, {Bravetti}, {Capozziello}, and {Luongo}}}]{aviles_2014}
\bibinfo{author}{\bibfnamefont{A.}~\bibnamefont{{Aviles}}}, \bibinfo{author}{\bibfnamefont{A.}~\bibnamefont{{Bravetti}}}, \bibinfo{author}{\bibfnamefont{S.}~\bibnamefont{{Capozziello}}}, \bibnamefont{and} \bibinfo{author}{\bibfnamefont{O.}~\bibnamefont{{Luongo}}}, \bibinfo{journal}{\prd} \textbf{\bibinfo{volume}{90}}, \bibinfo{eid}{043531} (\bibinfo{year}{2014}), \eprint{1405.6935}.

\bibitem[{\citenamefont{{Liu} and {Wei}}(2015)}]{jing_liu_2015}
\bibinfo{author}{\bibfnamefont{J.}~\bibnamefont{{Liu}}} \bibnamefont{and} \bibinfo{author}{\bibfnamefont{H.}~\bibnamefont{{Wei}}}, \bibinfo{journal}{General Relativity and Gravitation} \textbf{\bibinfo{volume}{47}}, \bibinfo{eid}{141} (\bibinfo{year}{2015}), \eprint{1410.3960}.

\bibitem[{\citenamefont{{Wei} et~al.}(2014)\citenamefont{{Wei}, {Yan}, and {Zhou}}}]{wei_2014}
\bibinfo{author}{\bibfnamefont{H.}~\bibnamefont{{Wei}}}, \bibinfo{author}{\bibfnamefont{X.-P.} \bibnamefont{{Yan}}}, \bibnamefont{and} \bibinfo{author}{\bibfnamefont{Y.-N.} \bibnamefont{{Zhou}}}, \bibinfo{journal}{\jcap} \textbf{\bibinfo{volume}{2014}}, \bibinfo{eid}{045} (\bibinfo{year}{2014}), \eprint{1312.1117}.

\bibitem[{\citenamefont{{Gruber} and {Luongo}}(2014)}]{gruber_2014}
\bibinfo{author}{\bibfnamefont{C.}~\bibnamefont{{Gruber}}} \bibnamefont{and} \bibinfo{author}{\bibfnamefont{O.}~\bibnamefont{{Luongo}}}, \bibinfo{journal}{\prd} \textbf{\bibinfo{volume}{89}}, \bibinfo{eid}{103506} (\bibinfo{year}{2014}), \eprint{1309.3215}.

\bibitem[{\citenamefont{{Adachi} and {Kasai}}(2012)}]{adachi_2012}
\bibinfo{author}{\bibfnamefont{M.}~\bibnamefont{{Adachi}}} \bibnamefont{and} \bibinfo{author}{\bibfnamefont{M.}~\bibnamefont{{Kasai}}}, \bibinfo{journal}{Progress of Theoretical Physics} \textbf{\bibinfo{volume}{127}}, \bibinfo{pages}{145} (\bibinfo{year}{2012}), \eprint{1111.6396}.

\bibitem[{\citenamefont{{Busti} et~al.}(2015)\citenamefont{{Busti}, {de la Cruz-Dombriz}, {Dunsby}, and {S{\'a}ez-G{\'o}mez}}}]{busti_2015}
\bibinfo{author}{\bibfnamefont{V.~C.} \bibnamefont{{Busti}}}, \bibinfo{author}{\bibfnamefont{{\'A}.}~\bibnamefont{{de la Cruz-Dombriz}}}, \bibinfo{author}{\bibfnamefont{P.~K.~S.} \bibnamefont{{Dunsby}}}, \bibnamefont{and} \bibinfo{author}{\bibfnamefont{D.}~\bibnamefont{{S{\'a}ez-G{\'o}mez}}}, \bibinfo{journal}{\prd} \textbf{\bibinfo{volume}{92}}, \bibinfo{eid}{123512} (\bibinfo{year}{2015}), \eprint{1505.05503}.

\bibitem[{\citenamefont{{Tucker} et~al.}(2005)\citenamefont{{Tucker}, {Burton}, and {Noble}}}]{tucker_2005}
\bibinfo{author}{\bibfnamefont{R.~W.} \bibnamefont{{Tucker}}}, \bibinfo{author}{\bibfnamefont{D.~A.} \bibnamefont{{Burton}}}, \bibnamefont{and} \bibinfo{author}{\bibfnamefont{A.}~\bibnamefont{{Noble}}}, \bibinfo{journal}{General Relativity and Gravitation} \textbf{\bibinfo{volume}{37}}, \bibinfo{pages}{1555} (\bibinfo{year}{2005}), \eprint{gr-qc/0411131}.

\bibitem[{\citenamefont{{Dunsby} and {Luongo}}(2016)}]{duncsby_2016}
\bibinfo{author}{\bibfnamefont{P.~K.~S.} \bibnamefont{{Dunsby}}} \bibnamefont{and} \bibinfo{author}{\bibfnamefont{O.}~\bibnamefont{{Luongo}}}, \bibinfo{journal}{International Journal of Geometric Methods in Modern Physics} \textbf{\bibinfo{volume}{13}}, \bibinfo{eid}{1630002-606} (\bibinfo{year}{2016}), \eprint{1511.06532}.

\bibitem[{\citenamefont{{Catto{\"e}n} and {Visser}}(2007)}]{catto_2007}
\bibinfo{author}{\bibfnamefont{C.}~\bibnamefont{{Catto{\"e}n}}} \bibnamefont{and} \bibinfo{author}{\bibfnamefont{M.}~\bibnamefont{{Visser}}}, \bibinfo{journal}{Classical and Quantum Gravity} \textbf{\bibinfo{volume}{24}}, \bibinfo{pages}{5985} (\bibinfo{year}{2007}), \eprint{0710.1887}.

\bibitem[{\citenamefont{{Capozziello} et~al.}(2020)\citenamefont{{Capozziello}, {D'Agostino}, and {Luongo}}}]{capozziello_2020}
\bibinfo{author}{\bibfnamefont{S.}~\bibnamefont{{Capozziello}}}, \bibinfo{author}{\bibfnamefont{R.}~\bibnamefont{{D'Agostino}}}, \bibnamefont{and} \bibinfo{author}{\bibfnamefont{O.}~\bibnamefont{{Luongo}}}, \bibinfo{journal}{\mnras} \textbf{\bibinfo{volume}{494}}, \bibinfo{pages}{2576} (\bibinfo{year}{2020}), \eprint{2003.09341}.

\bibitem[{\citenamefont{{Lobo} et~al.}(2020)\citenamefont{{Lobo}, {Mimoso}, and {Visser}}}]{lobo_2020}
\bibinfo{author}{\bibfnamefont{F.~S.~N.} \bibnamefont{{Lobo}}}, \bibinfo{author}{\bibfnamefont{J.~P.} \bibnamefont{{Mimoso}}}, \bibnamefont{and} \bibinfo{author}{\bibfnamefont{M.}~\bibnamefont{{Visser}}}, \bibinfo{journal}{\jcap} \textbf{\bibinfo{volume}{2020}}, \bibinfo{eid}{043} (\bibinfo{year}{2020}), \eprint{2001.11964}.

\bibitem[{\citenamefont{{mehrabi} and {Basilakos}}(2018)}]{mehrabi_2018}
\bibinfo{author}{\bibfnamefont{A.}~\bibnamefont{{mehrabi}}} \bibnamefont{and} \bibinfo{author}{\bibfnamefont{S.}~\bibnamefont{{Basilakos}}}, \bibinfo{journal}{European Physical Journal C} \textbf{\bibinfo{volume}{78}}, \bibinfo{eid}{889} (\bibinfo{year}{2018}), \eprint{1804.10794}.

\bibitem[{\citenamefont{{Zhou} et~al.}(2016{\natexlab{a}})\citenamefont{{Zhou}, {Liu}, {Zou}, and {Wei}}}]{zhou_2016a}
\bibinfo{author}{\bibfnamefont{Y.-N.} \bibnamefont{{Zhou}}}, \bibinfo{author}{\bibfnamefont{D.-Z.} \bibnamefont{{Liu}}}, \bibinfo{author}{\bibfnamefont{X.-B.} \bibnamefont{{Zou}}}, \bibnamefont{and} \bibinfo{author}{\bibfnamefont{H.}~\bibnamefont{{Wei}}}, \bibinfo{journal}{arXiv e-prints} \bibinfo{eid}{arXiv:1602.07189} (\bibinfo{year}{2016}{\natexlab{a}}), \eprint{1602.07189}.

\bibitem[{\citenamefont{{Zhou} et~al.}(2016{\natexlab{b}})\citenamefont{{Zhou}, {Liu}, {Zou}, and {Wei}}}]{zhou_2016b}
\bibinfo{author}{\bibfnamefont{Y.-N.} \bibnamefont{{Zhou}}}, \bibinfo{author}{\bibfnamefont{D.-Z.} \bibnamefont{{Liu}}}, \bibinfo{author}{\bibfnamefont{X.-B.} \bibnamefont{{Zou}}}, \bibnamefont{and} \bibinfo{author}{\bibfnamefont{H.}~\bibnamefont{{Wei}}}, \bibinfo{journal}{arXiv e-prints} \bibinfo{eid}{arXiv:1602.07189} (\bibinfo{year}{2016}{\natexlab{b}}), \eprint{1602.07189}.

\bibitem[{\citenamefont{{Liu} et~al.}(2021)\citenamefont{{Liu}, {Li}, {Yu}, and {Wu}}}]{yang_2021}
\bibinfo{author}{\bibfnamefont{Y.}~\bibnamefont{{Liu}}}, \bibinfo{author}{\bibfnamefont{Z.}~\bibnamefont{{Li}}}, \bibinfo{author}{\bibfnamefont{H.}~\bibnamefont{{Yu}}}, \bibnamefont{and} \bibinfo{author}{\bibfnamefont{P.}~\bibnamefont{{Wu}}}, \bibinfo{journal}{\apss} \textbf{\bibinfo{volume}{366}}, \bibinfo{eid}{112} (\bibinfo{year}{2021}), \eprint{2112.10959}.

\bibitem[{\citenamefont{{Capozziello} et~al.}(2018)\citenamefont{{Capozziello}, {D'Agostino}, and {Luongo}}}]{capozzielo_2018}
\bibinfo{author}{\bibfnamefont{S.}~\bibnamefont{{Capozziello}}}, \bibinfo{author}{\bibfnamefont{R.}~\bibnamefont{{D'Agostino}}}, \bibnamefont{and} \bibinfo{author}{\bibfnamefont{O.}~\bibnamefont{{Luongo}}}, \bibinfo{journal}{\jcap} \textbf{\bibinfo{volume}{2018}}, \bibinfo{eid}{008} (\bibinfo{year}{2018}), \eprint{1709.08407}.

\bibitem[{\citenamefont{{Dutta} et~al.}(2018)\citenamefont{{Dutta}, {Ruchika}, {Roy}, {Sen}, and {Sheikh-Jabbari}}}]{dutta_2018}
\bibinfo{author}{\bibfnamefont{K.}~\bibnamefont{{Dutta}}}, \bibinfo{author}{\bibnamefont{{Ruchika}}}, \bibinfo{author}{\bibfnamefont{A.}~\bibnamefont{{Roy}}}, \bibinfo{author}{\bibfnamefont{A.~A.} \bibnamefont{{Sen}}}, \bibnamefont{and} \bibinfo{author}{\bibfnamefont{M.~M.} \bibnamefont{{Sheikh-Jabbari}}}, \bibinfo{journal}{arXiv e-prints} \bibinfo{eid}{arXiv:1808.06623} (\bibinfo{year}{2018}), \eprint{1808.06623}.

\bibitem[{\citenamefont{{Dutta} et~al.}(2019)\citenamefont{{Dutta}, {Roy}, {Ruchika}, {Sen}, and {Sheikh-Jabbari}}}]{dutta_2019}
\bibinfo{author}{\bibfnamefont{K.}~\bibnamefont{{Dutta}}}, \bibinfo{author}{\bibfnamefont{A.}~\bibnamefont{{Roy}}}, \bibinfo{author}{\bibnamefont{{Ruchika}}}, \bibinfo{author}{\bibfnamefont{A.~A.} \bibnamefont{{Sen}}}, \bibnamefont{and} \bibinfo{author}{\bibfnamefont{M.~M.} \bibnamefont{{Sheikh-Jabbari}}}, \bibinfo{journal}{\prd} \textbf{\bibinfo{volume}{100}}, \bibinfo{eid}{103501} (\bibinfo{year}{2019}), \eprint{1908.07267}.

\bibitem[{\citenamefont{{D'Agostino} and {Nunes}}(2023)}]{agostino_2023}
\bibinfo{author}{\bibfnamefont{R.}~\bibnamefont{{D'Agostino}}} \bibnamefont{and} \bibinfo{author}{\bibfnamefont{R.~C.} \bibnamefont{{Nunes}}}, \bibinfo{journal}{\prd} \textbf{\bibinfo{volume}{108}}, \bibinfo{eid}{023523} (\bibinfo{year}{2023}), \eprint{2307.13464}.

\bibitem[{\citenamefont{{Scolnic} et~al.}(2022)\citenamefont{{Scolnic}, {Brout}, {Carr}, {Riess}, {Davis}, {Dwomoh}, {Jones}, {Ali}, {Charvu}, {Chen} et~al.}}]{scolnic}
\bibinfo{author}{\bibfnamefont{D.}~\bibnamefont{{Scolnic}}}, \bibinfo{author}{\bibfnamefont{D.}~\bibnamefont{{Brout}}}, \bibinfo{author}{\bibfnamefont{A.}~\bibnamefont{{Carr}}}, \bibinfo{author}{\bibfnamefont{A.~G.} \bibnamefont{{Riess}}}, \bibinfo{author}{\bibfnamefont{T.~M.} \bibnamefont{{Davis}}}, \bibinfo{author}{\bibfnamefont{A.}~\bibnamefont{{Dwomoh}}}, \bibinfo{author}{\bibfnamefont{D.~O.} \bibnamefont{{Jones}}}, \bibinfo{author}{\bibfnamefont{N.}~\bibnamefont{{Ali}}}, \bibinfo{author}{\bibfnamefont{P.}~\bibnamefont{{Charvu}}}, \bibinfo{author}{\bibfnamefont{R.}~\bibnamefont{{Chen}}}, \bibnamefont{et~al.}, \bibinfo{journal}{\apj} \textbf{\bibinfo{volume}{938}}, \bibinfo{eid}{113} (\bibinfo{year}{2022}), \eprint{2112.03863}.

\bibitem[{\citenamefont{{Jimenez} and {Loeb}}(2002)}]{Jimenez}
\bibinfo{author}{\bibfnamefont{R.}~\bibnamefont{{Jimenez}}} \bibnamefont{and} \bibinfo{author}{\bibfnamefont{A.}~\bibnamefont{{Loeb}}}, \bibinfo{journal}{\apj} \textbf{\bibinfo{volume}{573}}, \bibinfo{pages}{37} (\bibinfo{year}{2002}), \eprint{astro-ph/0106145}.

\bibitem[{\citenamefont{{Moresco} et~al.}(2012)\citenamefont{{Moresco}, {Cimatti}, {Jimenez}, {Pozzetti}, {Zamorani}, {Bolzonella}, {Dunlop}, {Lamareille}, {Mignoli}, {Pearce} et~al.}}]{moresco1}
\bibinfo{author}{\bibfnamefont{M.}~\bibnamefont{{Moresco}}}, \bibinfo{author}{\bibfnamefont{A.}~\bibnamefont{{Cimatti}}}, \bibinfo{author}{\bibfnamefont{R.}~\bibnamefont{{Jimenez}}}, \bibinfo{author}{\bibfnamefont{L.}~\bibnamefont{{Pozzetti}}}, \bibinfo{author}{\bibfnamefont{G.}~\bibnamefont{{Zamorani}}}, \bibinfo{author}{\bibfnamefont{M.}~\bibnamefont{{Bolzonella}}}, \bibinfo{author}{\bibfnamefont{J.}~\bibnamefont{{Dunlop}}}, \bibinfo{author}{\bibfnamefont{F.}~\bibnamefont{{Lamareille}}}, \bibinfo{author}{\bibfnamefont{M.}~\bibnamefont{{Mignoli}}}, \bibinfo{author}{\bibfnamefont{H.}~\bibnamefont{{Pearce}}}, \bibnamefont{et~al.}, \bibinfo{journal}{\jcap} \textbf{\bibinfo{volume}{2012}}, \bibinfo{eid}{006} (\bibinfo{year}{2012}), \eprint{1201.3609}.

\bibitem[{\citenamefont{{Moresco} et~al.}(2016)\citenamefont{{Moresco}, {Pozzetti}, {Cimatti}, {Jimenez}, {Maraston}, {Verde}, {Thomas}, {Citro}, {Tojeiro}, and {Wilkinson}}}]{moresco2}
\bibinfo{author}{\bibfnamefont{M.}~\bibnamefont{{Moresco}}}, \bibinfo{author}{\bibfnamefont{L.}~\bibnamefont{{Pozzetti}}}, \bibinfo{author}{\bibfnamefont{A.}~\bibnamefont{{Cimatti}}}, \bibinfo{author}{\bibfnamefont{R.}~\bibnamefont{{Jimenez}}}, \bibinfo{author}{\bibfnamefont{C.}~\bibnamefont{{Maraston}}}, \bibinfo{author}{\bibfnamefont{L.}~\bibnamefont{{Verde}}}, \bibinfo{author}{\bibfnamefont{D.}~\bibnamefont{{Thomas}}}, \bibinfo{author}{\bibfnamefont{A.}~\bibnamefont{{Citro}}}, \bibinfo{author}{\bibfnamefont{R.}~\bibnamefont{{Tojeiro}}}, \bibnamefont{and} \bibinfo{author}{\bibfnamefont{D.}~\bibnamefont{{Wilkinson}}}, \bibinfo{journal}{\jcap} \textbf{\bibinfo{volume}{2016}}, \bibinfo{eid}{014} (\bibinfo{year}{2016}), \eprint{1601.01701}.

\bibitem[{\citenamefont{{Moresco}}(2015)}]{moresco3}
\bibinfo{author}{\bibfnamefont{M.}~\bibnamefont{{Moresco}}}, \bibinfo{journal}{\mnras} \textbf{\bibinfo{volume}{450}}, \bibinfo{pages}{L16} (\bibinfo{year}{2015}), \eprint{1503.01116}.

\bibitem[{\citenamefont{{Ratsimbazafy} et~al.}(2017)\citenamefont{{Ratsimbazafy}, {Loubser}, {Crawford}, {Cress}, {Bassett}, {Nichol}, and {V{\"a}is{\"a}nen}}}]{ratsimbazafy}
\bibinfo{author}{\bibfnamefont{A.~L.} \bibnamefont{{Ratsimbazafy}}}, \bibinfo{author}{\bibfnamefont{S.~I.} \bibnamefont{{Loubser}}}, \bibinfo{author}{\bibfnamefont{S.~M.} \bibnamefont{{Crawford}}}, \bibinfo{author}{\bibfnamefont{C.~M.} \bibnamefont{{Cress}}}, \bibinfo{author}{\bibfnamefont{B.~A.} \bibnamefont{{Bassett}}}, \bibinfo{author}{\bibfnamefont{R.~C.} \bibnamefont{{Nichol}}}, \bibnamefont{and} \bibinfo{author}{\bibfnamefont{P.}~\bibnamefont{{V{\"a}is{\"a}nen}}}, \bibinfo{journal}{\mnras} \textbf{\bibinfo{volume}{467}}, \bibinfo{pages}{3239} (\bibinfo{year}{2017}), \eprint{1702.00418}.

\bibitem[{\citenamefont{{Stern} et~al.}(2010)\citenamefont{{Stern}, {Jimenez}, {Verde}, {Kamionkowski}, and {Stanford}}}]{stern}
\bibinfo{author}{\bibfnamefont{D.}~\bibnamefont{{Stern}}}, \bibinfo{author}{\bibfnamefont{R.}~\bibnamefont{{Jimenez}}}, \bibinfo{author}{\bibfnamefont{L.}~\bibnamefont{{Verde}}}, \bibinfo{author}{\bibfnamefont{M.}~\bibnamefont{{Kamionkowski}}}, \bibnamefont{and} \bibinfo{author}{\bibfnamefont{S.~A.} \bibnamefont{{Stanford}}}, \bibinfo{journal}{\jcap} \textbf{\bibinfo{volume}{2010}}, \bibinfo{eid}{008} (\bibinfo{year}{2010}), \eprint{0907.3149}.

\bibitem[{\citenamefont{{Borghi} et~al.}(2022)\citenamefont{{Borghi}, {Moresco}, and {Cimatti}}}]{borghi}
\bibinfo{author}{\bibfnamefont{N.}~\bibnamefont{{Borghi}}}, \bibinfo{author}{\bibfnamefont{M.}~\bibnamefont{{Moresco}}}, \bibnamefont{and} \bibinfo{author}{\bibfnamefont{A.}~\bibnamefont{{Cimatti}}}, \bibinfo{journal}{\apjl} \textbf{\bibinfo{volume}{928}}, \bibinfo{eid}{L4} (\bibinfo{year}{2022}), \eprint{2110.04304}.

\bibitem[{\citenamefont{{Simon} et~al.}(2005)\citenamefont{{Simon}, {Verde}, and {Jimenez}}}]{joan}
\bibinfo{author}{\bibfnamefont{J.}~\bibnamefont{{Simon}}}, \bibinfo{author}{\bibfnamefont{L.}~\bibnamefont{{Verde}}}, \bibnamefont{and} \bibinfo{author}{\bibfnamefont{R.}~\bibnamefont{{Jimenez}}}, \bibinfo{journal}{\prd} \textbf{\bibinfo{volume}{71}}, \bibinfo{eid}{123001} (\bibinfo{year}{2005}), \eprint{astro-ph/0412269}.

\bibitem[{\citenamefont{{Zhang} et~al.}(2014)\citenamefont{{Zhang}, {Zhang}, {Yuan}, {Liu}, {Zhang}, and {Sun}}}]{cong}
\bibinfo{author}{\bibfnamefont{C.}~\bibnamefont{{Zhang}}}, \bibinfo{author}{\bibfnamefont{H.}~\bibnamefont{{Zhang}}}, \bibinfo{author}{\bibfnamefont{S.}~\bibnamefont{{Yuan}}}, \bibinfo{author}{\bibfnamefont{S.}~\bibnamefont{{Liu}}}, \bibinfo{author}{\bibfnamefont{T.-J.} \bibnamefont{{Zhang}}}, \bibnamefont{and} \bibinfo{author}{\bibfnamefont{Y.-C.} \bibnamefont{{Sun}}}, \bibinfo{journal}{Research in Astronomy and Astrophysics} \textbf{\bibinfo{volume}{14}}, \bibinfo{eid}{1221-1233} (\bibinfo{year}{2014}), \eprint{1207.4541}.

\bibitem[{\citenamefont{{Moresco} et~al.}(2020)\citenamefont{{Moresco}, {Jimenez}, {Verde}, {Cimatti}, and {Pozzetti}}}]{moresco4}
\bibinfo{author}{\bibfnamefont{M.}~\bibnamefont{{Moresco}}}, \bibinfo{author}{\bibfnamefont{R.}~\bibnamefont{{Jimenez}}}, \bibinfo{author}{\bibfnamefont{L.}~\bibnamefont{{Verde}}}, \bibinfo{author}{\bibfnamefont{A.}~\bibnamefont{{Cimatti}}}, \bibnamefont{and} \bibinfo{author}{\bibfnamefont{L.}~\bibnamefont{{Pozzetti}}}, \bibinfo{journal}{\apj} \textbf{\bibinfo{volume}{898}}, \bibinfo{eid}{82} (\bibinfo{year}{2020}), \eprint{2003.07362}.

\bibitem[{\citenamefont{Moresco}(2023)}]{Moresco_2307}
\bibinfo{author}{\bibfnamefont{M.}~\bibnamefont{Moresco}} (\bibinfo{year}{2023}), \eprint{2307.09501}.

\bibitem[{\citenamefont{{G{\'o}mez-Valent}}(2022)}]{36}
\bibinfo{author}{\bibfnamefont{A.}~\bibnamefont{{G{\'o}mez-Valent}}}, \bibinfo{journal}{\prd} \textbf{\bibinfo{volume}{105}}, \bibinfo{eid}{043528} (\bibinfo{year}{2022}), \eprint{2111.15450}.

\bibitem[{\citenamefont{{Seikel} et~al.}(2012)\citenamefont{{Seikel}, {Clarkson}, and {Smith}}}]{Seikel}
\bibinfo{author}{\bibfnamefont{M.}~\bibnamefont{{Seikel}}}, \bibinfo{author}{\bibfnamefont{C.}~\bibnamefont{{Clarkson}}}, \bibnamefont{and} \bibinfo{author}{\bibfnamefont{M.}~\bibnamefont{{Smith}}}, \bibinfo{journal}{\jcap} \textbf{\bibinfo{volume}{2012}}, \bibinfo{eid}{036} (\bibinfo{year}{2012}), \eprint{1204.2832}.

\bibitem[{\citenamefont{{Dinda} and {Banerjee}}(2023)}]{Dinda}
\bibinfo{author}{\bibfnamefont{B.~R.} \bibnamefont{{Dinda}}} \bibnamefont{and} \bibinfo{author}{\bibfnamefont{N.}~\bibnamefont{{Banerjee}}}, \bibinfo{journal}{\prd} \textbf{\bibinfo{volume}{107}}, \bibinfo{eid}{063513} (\bibinfo{year}{2023}), \eprint{2208.14740}.

\bibitem[{\citenamefont{{Camarena} and {Marra}}(2020)}]{7}
\bibinfo{author}{\bibfnamefont{D.}~\bibnamefont{{Camarena}}} \bibnamefont{and} \bibinfo{author}{\bibfnamefont{V.}~\bibnamefont{{Marra}}}, \bibinfo{journal}{\mnras} \textbf{\bibinfo{volume}{495}}, \bibinfo{pages}{2630} (\bibinfo{year}{2020}), \eprint{1910.14125}.

\bibitem[{\citenamefont{{Greene} and {Cyr-Racine}}(2022)}]{28Greene}
\bibinfo{author}{\bibfnamefont{K.~L.} \bibnamefont{{Greene}}} \bibnamefont{and} \bibinfo{author}{\bibfnamefont{F.-Y.} \bibnamefont{{Cyr-Racine}}}, \bibinfo{journal}{\jcap} \textbf{\bibinfo{volume}{2022}}, \bibinfo{eid}{002} (\bibinfo{year}{2022}), \eprint{2112.11567}.

\bibitem[{\citenamefont{{Wojtak} and {Agnello}}(2019)}]{wojtak}
\bibinfo{author}{\bibfnamefont{R.}~\bibnamefont{{Wojtak}}} \bibnamefont{and} \bibinfo{author}{\bibfnamefont{A.}~\bibnamefont{{Agnello}}}, \bibinfo{journal}{\mnras} \textbf{\bibinfo{volume}{486}}, \bibinfo{pages}{5046} (\bibinfo{year}{2019}), \eprint{1908.02401}.

\bibitem[{\citenamefont{{Liu} et~al.}(2025)\citenamefont{{Liu}, {Cao}, and {Wang}}}]{liusoundhorizon}
\bibinfo{author}{\bibfnamefont{T.}~\bibnamefont{{Liu}}}, \bibinfo{author}{\bibfnamefont{S.}~\bibnamefont{{Cao}}}, \bibnamefont{and} \bibinfo{author}{\bibfnamefont{J.}~\bibnamefont{{Wang}}}, \bibinfo{journal}{\prd} \textbf{\bibinfo{volume}{111}}, \bibinfo{eid}{023524} (\bibinfo{year}{2025}), \eprint{2406.18298}.

\bibitem[{\citenamefont{{Zhang} and {Huang}}(2021)}]{zhang}
\bibinfo{author}{\bibfnamefont{X.}~\bibnamefont{{Zhang}}} \bibnamefont{and} \bibinfo{author}{\bibfnamefont{Q.-G.} \bibnamefont{{Huang}}}, \bibinfo{journal}{\prd} \textbf{\bibinfo{volume}{103}}, \bibinfo{eid}{043513} (\bibinfo{year}{2021}), \eprint{2006.16692}.

\bibitem[{\citenamefont{{Lange}}(2023)}]{nautilus}
\bibinfo{author}{\bibfnamefont{J.~U.} \bibnamefont{{Lange}}}, \bibinfo{journal}{\mnras} \textbf{\bibinfo{volume}{525}}, \bibinfo{pages}{3181} (\bibinfo{year}{2023}), \eprint{2306.16923}.

\bibitem[{\citenamefont{{Lewis}}(2019)}]{getdist}
\bibinfo{author}{\bibfnamefont{A.}~\bibnamefont{{Lewis}}}, \bibinfo{journal}{arXiv e-prints} \bibinfo{eid}{arXiv:1910.13970} (\bibinfo{year}{2019}), \eprint{1910.13970}.

\bibitem[{\citenamefont{{Colg{\'a}in} et~al.}(2024)\citenamefont{{Colg{\'a}in}, {Dainotti}, {Capozziello}, {Pourojaghi}, {Sheikh-Jabbari}, and {Stojkovic}}}]{colgain}
\bibinfo{author}{\bibfnamefont{E.~{\'O}.} \bibnamefont{{Colg{\'a}in}}}, \bibinfo{author}{\bibfnamefont{M.~G.} \bibnamefont{{Dainotti}}}, \bibinfo{author}{\bibfnamefont{S.}~\bibnamefont{{Capozziello}}}, \bibinfo{author}{\bibfnamefont{S.}~\bibnamefont{{Pourojaghi}}}, \bibinfo{author}{\bibfnamefont{M.~M.} \bibnamefont{{Sheikh-Jabbari}}}, \bibnamefont{and} \bibinfo{author}{\bibfnamefont{D.}~\bibnamefont{{Stojkovic}}}, \bibinfo{journal}{arXiv e-prints} \bibinfo{eid}{arXiv:2404.08633} (\bibinfo{year}{2024}), \eprint{2404.08633}.

\bibitem[{\citenamefont{{Patel} et~al.}(2024)\citenamefont{{Patel}, {Chakraborty}, and {Amendola}}}]{Patel}
\bibinfo{author}{\bibfnamefont{V.}~\bibnamefont{{Patel}}}, \bibinfo{author}{\bibfnamefont{A.}~\bibnamefont{{Chakraborty}}}, \bibnamefont{and} \bibinfo{author}{\bibfnamefont{L.}~\bibnamefont{{Amendola}}}, \bibinfo{journal}{arXiv e-prints} \bibinfo{eid}{arXiv:2407.06586} (\bibinfo{year}{2024}), \eprint{2407.06586}.

\bibitem[{\citenamefont{{Chen} et~al.}(2024)\citenamefont{{Chen}, {Kumar}, {Ratra}, and {Xu}}}]{Chen_2024}
\bibinfo{author}{\bibfnamefont{Y.}~\bibnamefont{{Chen}}}, \bibinfo{author}{\bibfnamefont{S.}~\bibnamefont{{Kumar}}}, \bibinfo{author}{\bibfnamefont{B.}~\bibnamefont{{Ratra}}}, \bibnamefont{and} \bibinfo{author}{\bibfnamefont{T.}~\bibnamefont{{Xu}}}, \bibinfo{journal}{\apjl} \textbf{\bibinfo{volume}{964}}, \bibinfo{eid}{L4} (\bibinfo{year}{2024}), \eprint{2401.13187}.

\bibitem[{\citenamefont{{Brout} et~al.}(2022)\citenamefont{{Brout}, {Scolnic}, {Popovic}, {Riess}, {Carr}, {Zuntz}, {Kessler}, {Davis}, {Hinton}, {Jones} et~al.}}]{pantheon}
\bibinfo{author}{\bibfnamefont{D.}~\bibnamefont{{Brout}}}, \bibinfo{author}{\bibfnamefont{D.}~\bibnamefont{{Scolnic}}}, \bibinfo{author}{\bibfnamefont{B.}~\bibnamefont{{Popovic}}}, \bibinfo{author}{\bibfnamefont{A.~G.} \bibnamefont{{Riess}}}, \bibinfo{author}{\bibfnamefont{A.}~\bibnamefont{{Carr}}}, \bibinfo{author}{\bibfnamefont{J.}~\bibnamefont{{Zuntz}}}, \bibinfo{author}{\bibfnamefont{R.}~\bibnamefont{{Kessler}}}, \bibinfo{author}{\bibfnamefont{T.~M.} \bibnamefont{{Davis}}}, \bibinfo{author}{\bibfnamefont{S.}~\bibnamefont{{Hinton}}}, \bibinfo{author}{\bibfnamefont{D.}~\bibnamefont{{Jones}}}, \bibnamefont{et~al.}, \bibinfo{journal}{\apj} \textbf{\bibinfo{volume}{938}}, \bibinfo{eid}{110} (\bibinfo{year}{2022}), \eprint{2202.04077}.

\bibitem[{\citenamefont{{Moresco}}(2023)}]{moresco}
\bibinfo{author}{\bibfnamefont{M.}~\bibnamefont{{Moresco}}}, \bibinfo{journal}{arXiv e-prints} \bibinfo{eid}{arXiv:2307.09501} (\bibinfo{year}{2023}), \eprint{2307.09501}.

\bibitem[{\citenamefont{{Ruchika}}(2024)}]{ruchika_2024}
\bibinfo{author}{\bibnamefont{{Ruchika}}}, \bibinfo{journal}{arXiv e-prints} \bibinfo{eid}{arXiv:2406.05453} (\bibinfo{year}{2024}), \eprint{2406.05453}.

\bibitem[{\citenamefont{{Trotta}}(2017)}]{trotta_2017}
\bibinfo{author}{\bibfnamefont{R.}~\bibnamefont{{Trotta}}}, \bibinfo{journal}{arXiv e-prints} \bibinfo{eid}{arXiv:1701.01467} (\bibinfo{year}{2017}), \eprint{1701.01467}.

\bibitem[{\citenamefont{{Krishak} and {Desai}}(2020)}]{Krishak20}
\bibinfo{author}{\bibfnamefont{A.}~\bibnamefont{{Krishak}}} \bibnamefont{and} \bibinfo{author}{\bibfnamefont{S.}~\bibnamefont{{Desai}}}, \bibinfo{journal}{\jcap} \textbf{\bibinfo{volume}{2020}}, \bibinfo{eid}{006} (\bibinfo{year}{2020}), \eprint{2003.10127}.

\bibitem[{\citenamefont{Cao and Ratra}(2023)}]{caoratra}
\bibinfo{author}{\bibfnamefont{S.}~\bibnamefont{Cao}} \bibnamefont{and} \bibinfo{author}{\bibfnamefont{B.}~\bibnamefont{Ratra}}, \bibinfo{journal}{Phys. Rev. D} \textbf{\bibinfo{volume}{107}}, \bibinfo{pages}{103521} (\bibinfo{year}{2023}), \eprint{2302.14203}.

\bibitem[{\citenamefont{Freedman et~al.}(2024)\citenamefont{Freedman, Madore, Jang, Hoyt, Lee, and Owens}}]{Freedman_2024}
\bibinfo{author}{\bibfnamefont{W.~L.} \bibnamefont{Freedman}}, \bibinfo{author}{\bibfnamefont{B.~F.} \bibnamefont{Madore}}, \bibinfo{author}{\bibfnamefont{I.~S.} \bibnamefont{Jang}}, \bibinfo{author}{\bibfnamefont{T.~J.} \bibnamefont{Hoyt}}, \bibinfo{author}{\bibfnamefont{A.~J.} \bibnamefont{Lee}}, \bibnamefont{and} \bibinfo{author}{\bibfnamefont{K.~A.} \bibnamefont{Owens}} (\bibinfo{year}{2024}), \eprint{2408.06153}.

\bibitem[{\citenamefont{{Hern{\'a}ndez-Almada} et~al.}(2024)\citenamefont{{Hern{\'a}ndez-Almada}, {Mendoza-Mart{\'\i}nez}, {Garc{\'\i}a-Aspeitia}, and {Motta}}}]{recentdesi}
\bibinfo{author}{\bibfnamefont{A.}~\bibnamefont{{Hern{\'a}ndez-Almada}}}, \bibinfo{author}{\bibfnamefont{M.~L.} \bibnamefont{{Mendoza-Mart{\'\i}nez}}}, \bibinfo{author}{\bibfnamefont{M.~A.} \bibnamefont{{Garc{\'\i}a-Aspeitia}}}, \bibnamefont{and} \bibinfo{author}{\bibfnamefont{V.}~\bibnamefont{{Motta}}}, \bibinfo{journal}{Physics of the Dark Universe} \textbf{\bibinfo{volume}{46}}, \bibinfo{eid}{101668} (\bibinfo{year}{2024}), \eprint{2412.13045}.

\bibitem[{\citenamefont{{Benisty} et~al.}(2023)\citenamefont{{Benisty}, {Mifsud}, {Levi Said}, and {Staicova}}}]{benitsy}
\bibinfo{author}{\bibfnamefont{D.}~\bibnamefont{{Benisty}}}, \bibinfo{author}{\bibfnamefont{J.}~\bibnamefont{{Mifsud}}}, \bibinfo{author}{\bibfnamefont{J.}~\bibnamefont{{Levi Said}}}, \bibnamefont{and} \bibinfo{author}{\bibfnamefont{D.}~\bibnamefont{{Staicova}}}, \bibinfo{journal}{Physics of the Dark Universe} \textbf{\bibinfo{volume}{39}}, \bibinfo{eid}{101160} (\bibinfo{year}{2023}), \eprint{2202.04677}.

\end{thebibliography}

\appendix

\section{Constraints using Pantheon$+$ and DESI-BAO data only}
\label{appA}

Here, we present the results of the analysis done without using Cosmic Chronometer dataset (Tables \ref{table10}, \ref{table11}, \ref{table12}, \ref{table13}). As will be noticed, in the case of applying uniform priors on both $M$ and $r_d$, the estimated values of $H_0$ have larger errors without CC. Due to this, we included CC data in the main analysis. However, the results are similar to the ones obtained using the dataset combination CC$+$Pantheon Plus$+$BAO. 

\begin{table}[htbp!]
\caption{\rthis{Discrepancy} in $H_0$ estimate compared to that from Planck Cosmology~\cite{Planck} for a uniform prior on $M$ and a uniform prior on $r_d \in (0, 200)$. The standard prior combination (cf. Table~\ref{table2}) in Table \ref{table13} has Bayes' Factor of 1. The corresponding table which includes CC data for the same priors is Table~\ref{table2}.}
\label{table10}
\centering
    \begin{tabular}{|c|c|c|c|c|c|}
        \hline
        \thead{\boldmath$M$} & \thead{\boldmath$r_d$ (Mpc)} & \thead{\boldmath$H_0$ (km/s/Mpc)} & \thead{\boldmath$\Omega_m$} & \thead{Bayes' Factor} & \thead{Discrepancy (in $\sigma)$}\\
        \hline
        $-19.04\pm0.6$ & $129.3^{+20}_{-50}$ & $85^{+10}_{-30}$ & $0.305\pm0.013$ & 354 & 0.89 \\
        \hline
    \end{tabular}
\end{table}

\begin{table}[htbp!]
\caption{\rthis{Discrepancy} in $H_0$ estimate compared to that from Planck Cosmology~\cite{Planck} for a Gaussian prior on $M$ and a uniform prior on $r_d \in (0, 200)$. The standard prior combination (cf. Table~\ref{table2}) in Table \ref{table13} has Bayes' Factor of 1. The corresponding table which includes CC data  for the same priors is Table~\ref{table4}.}
\label{table11}
\centering
    \begin{tabular}{|c|c|c|c|c|c|}
        \hline
        \thead{\boldmath$M$} & \thead{\boldmath$r_d$ (Mpc)} & \thead{\boldmath$H_0$ (km/s/Mpc)} & \thead{\boldmath$\Omega_m$} & \thead{Bayes' Factor} & \thead{Discrepancy (in $\sigma)$}\\
        \hline
        $-19.253\pm0.027$ & $137.2\pm2.1$ & $73.72\pm0.95$ & $0.305\pm0.013$ &  354 & 5.82 \\
        $-19.362\pm0.072$ & $144.3\pm4.8$ & $70.1\pm2.3$   & $0.305\pm0.013$ &  804 & 1.16 \\
        $-19.396\pm0.015$ & $146.5\pm1.7$ & $69.03\pm0.56$ & $0.305\pm0.013$ & 1012 & 2.15 \\
        $-19.401\pm0.027$ & $146.8\pm2.2$ & $68.89\pm0.90$ & $0.305\pm0.013$ & 972 & 1.46 \\
        $-19.420\pm0.014$ & $148.1\pm1.6$ & $68.27\pm0.52$ & $0.305\pm0.013$ &  880 & 1.21 \\
        \hline
    \end{tabular}
\end{table}

\begin{table}[htbp!]
\caption{\rthis{Discrepancy} in $H_0$ estimate compared to that from Planck Cosmology~\cite{Planck} for a Gaussian prior on $r_d$ and uniform prior on $M \in (-21, -18)$. The standard prior combination (cf. Table~\ref{table2}) in Table \ref{table13}  has Bayes' Factor of 1. The corresponding table which includes CC data  for the same priors is Table~\ref{table6}.}
\label{table12}
\centering
    \begin{tabular}{|c|c|c|c|c|c|}
        \hline
        \thead{\boldmath$r_d$ (Mpc)} & \thead{\boldmath$M$} & \thead{\boldmath$H_0$ (km/s/Mpc)} & \thead{\boldmath$\Omega_m$} & \thead{Bayes' Factor} & \thead{Discrepancy (in $\sigma)$}\\
        \hline
        $137.00\pm4.5$  & $-19.248\pm0.073$ & $73.9\pm2.5$   & $0.305\pm0.013$  & 317 & 2.56 \\
        $139.70\pm4.85$ & $-19.290\pm0.077$ & $72.5\pm2.6$   & $0.305\pm0.013$  & 464 & 1.94 \\
        $147.05\pm0.3$  & $-19.405\pm0.02$  & $68.75\pm0.77$ & $0.305\pm0.013$  & 706 & 1.48 \\
        $148.00\pm3.6$  & $-19.417\pm0.056$ & $68.4\pm1.8$   & $0.305\pm0.013$ & 590 & 0.55 \\
        \hline
    \end{tabular}
\end{table}

\begin{table}[htbp!]
\caption{\rthis{Discrepancy} in $H_0$ estimate compared to that from Planck Cosmology~\cite{Planck}  for a  Gaussian prior on $r_d$ and $M$. For the standard prior combination (cf. Table~\ref{table2}), the Bayes' Factor is 1 which is the null hypothesis. The corresponding table which includes CC data for the same priors  is Table~\ref{table8}.}
\label{table13}
\centering
    \begin{tabular}{|c|c|c|c|c|c|}
        \hline
        \thead{\boldmath$M$ prior} & \thead{\boldmath$r_d$ (Mpc)} & \thead{\boldmath$H_0$ (km/s/Mpc)} & \thead{\boldmath$\Omega_m$} & \thead{Bayes' Factor} & \thead{Discrepancy (in $\sigma)$}\\
        \hline
        \multirow{4}{8em}{$-19.253\pm0.027$} & $137\pm4.5$ & $73.76\pm0.90$ & $0.305\pm0.013$ & $4.2\times10^3$ & 6.1 \\
        & $139.7\pm4.85$ & $73.59\pm0.92$ & $0.304\pm0.013$ & $3.9\times10^3$ & 5.84 \\
        & $147.05\pm0.3$ & $70.70\pm0.64$ & $0.279\pm0.011$ & 1 & 3.98 \\
        & $148\pm3.6$    & $72.84\pm0.87$ & $0.297\pm0.012$ & 320 & 5.35 \\
        \hline
        \multirow{4}{8em}{$-19.362\pm0.072$} & $137\pm4.5$ & $71.9\pm1.7$ & $0.307\pm0.013$ & $3.5\times10^3$ & 2.54 \\
        & $139.7\pm4.85$ & $71.2\pm1.7$   & $0.306\pm0.013$ & $5.8\times10^3$ & 2.15 \\
        & $147.05\pm0.3$ & $68.86\pm0.74$ & $0.304\pm0.013$ & $10^4$ & 1.64 \\
        & $148\pm3.6$    & $69.0\pm1.5$   & $0.304\pm0.013$ & $7.3\times10^3$ & 1.03 \\
        \hline
        \multirow{4}{8em}{$-19.396\pm0.015$} & $137\pm4.5$ & $69.19\pm0.56$ & $0.311\pm0.013$ & $1.9\times10^3$ & 2.35 \\
        & $139.7\pm4.85$  & $69.13\pm0.56$ & $0.309\pm0.013$ & $5.1\times10^3$ & 2.28 \\
        & $147.05\pm0.3$  & $68.94\pm0.51$ & $0.302\pm0.010$ & $33\times10^3$ & 2.13 \\
        & $148\pm3.6$     & $68.99\pm0.55$ & $0.304\pm0.012$ & $13.9\times10^3$ & 2.12 \\
        \hline
        \multirow{4}{8em}{$-19.401\pm0.027$} & $137\pm4.5$ & $69.44\pm0.85$ & $0.310\pm0.013$ & $1.8\times10^3$ & 2.07 \\
        & $139.7\pm4.85$ & $69.24\pm0.85$ & $0.309\pm0.013$ & $4.8\times10^3$ & 1.87 \\
        & $147.05\pm0.3$ & $68.80\pm0.64$ & $0.304\pm0.011$ & $26\times10^3$ & 1.72 \\
        & $148\pm3.6$    & $68.76\pm0.84$ & $0.304\pm0.013$ & $12.8\times10^3$ & 1.41 \\
        \hline
        \multirow{4}{8em}{$-19.420\pm0.014$} & $137\pm4.5$ & $68.42\pm0.52$ & $0.312\pm0.012$ & $837$ & 1.41 \\
        & $139.7\pm4.85$ & $68.37\pm0.52$ & $0.310\pm0.013$ & $2.9\times10^3$ & 1.35 \\
        & $147.05\pm0.3$ & $68.39\pm0.50$ & $0.31\pm0.01$ & $29\times10^3$ & 1.4 \\
        & $148\pm3.6$    & $68.27\pm0.52$ & $0.305\pm0.013$ & $13.5\times10^3$ & 1.21 \\
        \hline
    \end{tabular}
\end{table}

\end{document}